\documentclass[twocolumn,twocolappendix]{aastex63}
\usepackage{amsmath}

\newcommand\kms{km s$^{-1}$}
\newcommand\masyr{mas yr$^{-1}$}

\newcommand\msun{M$_\odot$}
\newcommand\rsun{R$_\odot$}

\defcitealias{kounkel2018a}{Paper I}
\usepackage{listings}

\begin{document}
\title{Dynamical star forming history of Per OB2}

\author[0000-0002-5365-1267]{Marina Kounkel}
\affil{Department of Physics and Astronomy, Vanderbilt University, VU Station 1807, Nashville, TN 37235, USA}
\email{marina.kounkel@vanderbilt.edu}
\author{TingYan Deng}
\affil{Department of Electrical Engineering and Computer Science, Vanderbilt University, VU Station 1824, Nashville, TN 37235, USA}
\author[0000-0002-3481-9052]{Keivan G.\ Stassun}
\affil{Department of Physics and Astronomy, Vanderbilt University, VU Station 1807, Nashville, TN 37235, USA} 

\begin{abstract}

We analyze the internal dynamics of young stars towards Perseus using Gaia EDR3 data, including Per OB2 and California Cloud. Interpreting the current dynamics, we speculate that Per OB2 may have formed from two separate clouds that have begun forming stars in a close proximity of each other. IC 348 is caught in the middle between the two of them, inheriting kinematics of both, and it stands out as a possible site of cloud cloud interaction. We also consider a possibility of a past supernova in Per OB2 - while one had likely occurred, it did not appear to have caused any obvious triggered star formation, but it has created a shock which has swept away the molecular gas away from IC 348. Finally, we examine a recently proposed shell between Taurus and Perseus. While its origin is unknown, we find no support for an expanding bubble in stellar kinematics, nor can we identify a likely progenitor for a supernova that may have caused it, disfavoring this scenario in the formation of this apparent shell.
\end{abstract}

\keywords{}

\section{Introduction}

Analysis of dynamical evolution of various star forming regions is becoming an increasing focus of studies with the release of astrometry from Gaia \citep{gaia-collaboration2021}. Previously, several regions have received much focus, like the Orion complex \citep[e.g.,][]{kounkel2020,chen2020a,swiggum2021}, Taurus \citep[e.g.,][]{krolikowski2021}, or Sco-Cen \citep[e.g.,][]{damiani2019,kerr2021,luhman2022a}. A particular interest of such studies is characterizing internal dynamics of an association, and how these dynamics correlate with the age of stars within it to understand how star formation propagates.

In this paper we examine Per OB2 association, which has a typical age of 2--5 Myr \citep{bally2008a}. This association has been classically defined to be located between $156<l<165^\circ$ and $-21<b<-11^\circ$ \citep{blaauw1964}. However, historically, the bulk of the analysis of this region has been primarily focused only on a small part of Per OB2, namely, the two younger clusters within it, NGC 1333 and IC 348 \citep[e.g.,][]{stelzer2012,luhman2016,hacar2017}. These clusters stand out as they are still associated with the molecular gas, and contain a large fraction of protostars, as well as young stars with disks, making it possible to identify them through various surveys, e.g., with Spitzer \citep{werner2006,gehrz2007} and others, and analyze their distribution within each cluster, as well as cluster properties \citep[e.g.,][]{muench2007,gutermuth2008}. On the other hand, other parts of Per OB2, being somewhat older (and thus its members more difficult to identify prior to Gaia) have not received comparable scrutiny.

Recently, \citet{bialy2021} have examined the 3-dimensional structure of the molecular gas and dust towards Perseus, using the 3D dust map from \citet{leike2020}. They proposed a possibility of a bubble between Taurus and Perseus molecular clouds that created the shell of gas from an event 6--22 Myr ago. However, as they focus only on the gas, they similarly exclude older populations from consideration, the distribution and the kinematics of which may have implications on the observed ``shell''.

There have been early explorations of the full morphology of Per OB2 with Hipparcos \citep{belikov2002,bouy2015}, but it took until the advent of Gaia to fully delve into its internal dynamics \citep{kounkel2019a,pavlidou2021}. These studies revealed a significant complexity in its structure. In particular, \citet{pavlidou2021} have identified 7 distinct groups into which Per OB2 can be split, of which 5 groups were newly defined. They also found that these groups can be separated into two sets, based on their dominant proper motions. But, the dynamical analysis was primarily conducted in the plane of the sky, which made it difficult to fully disentangle its inner workings. A full 3d analysis is needed to better understand the initial conditions and the star forming history of this region to understand the processes that have led to the current day dynamical state of Per OB2. As such, many of the questions posed in the review of the region \citep{bally2008a} still remain.

We analyze the internal structure and dynamics of Per OB2. In Section \ref{sec:data} we discuss the data we use for this analysis. In Section \ref{sec:method} we present the methodology of how these data has been processed to create a coarse map of these populations via clustering. In Section \ref{sec:results}  we discuss the identified structures. In Section \ref{sec:discussion} we discuss the implications on star forming history of Per OB2. Finally in Section \ref{sec:concl} we offer our conclusions.

\section{Data}\label{sec:data}

We have used Gaia EDR3 data \citep{gaia-collaboration2021} to obtain a census of stars towards Perseus. We require:

\begin{itemize}
\item $150<l<175^\circ$, $-25<b<-5^\circ$
\item $0<\mu_\alpha<10$ \masyr, $-14<\mu_\delta<-2$ \masyr
\item$1.5<\pi<5$ mas.
\item G$<18$ mag
\item RUWE$<1.5$
\end{itemize}

The structure of Per OB2 has been previously studied by \citet{pavlidou2021}, however the search radius was somewhat limited. This selection extends the analyzed area on the sky to include the California cloud \citep{lada2009} which \citet{kounkel2019a} have suggested to be associated with Per OB2.

The resulting catalog consisted of 40,737 stars towards Perseus.

In addition to the astrometry, we use RVs from Gaia \citep{cropper2018,gaia-collaboration2018,gaia-collaboration2021}. We have also supplemented in RVs from medium resolution LAMOST DR7 \citep{luo2019} and from APOGEE DR17 \citep{abdurrouf2022} when possible. 

\section{Methodology}\label{sec:method}

\subsection{Clustering}

Previously \citet{kounkel2019a} and \citet{kounkel2020} have performed hierarchical clustering across the entire sky with HDBSCAN \citep{hdbscan1,hdbscan}. As such, while the overall morphology of the underlying structure is known, the goal of those papers was to identify the largest coherent structure associated with each population, with all the substructures merged together. Instead, we now fine-tune the HDBSCAN for Perseus to better identify distinct smaller scale substructure within each region.

As such, we perform clustering on the Gaia data described in Section \ref{sec:data}. This identifies young populations, and separates the likely members from the field contaminants.

Clustering was performed in 5 dimensions, $l$, $b$, $\pi$, as well as proper motions. To avoid distortions in proper motions along the sky, they were converted to the local standard of rest \citep[using coefficients from][]{schonrich2010}; furthermore, they were converted from proper motions in \masyr\ to tangential velocities $v_{l,lsr}$ and $v_{b,lsr}$ in \kms\ to avoid distortion in distance\footnote{See \url{https://colab.research.google.com/drive/1Zf5PpwEBdWjfspsNKR0ZCDn5MM3EM0Ta} for the example code of the coordinate transformation and clustering.}. While the precise conversion to the LSR is only accurate to $\sim$1 \kms\ across the entire sky, in smaller spatial scales this uncertainty reduces to a largely uniform zero-point offset while allowing to correct for a much larger apparent gradient, as such applying this correction is more advantageous rather than not. Various additional normalization scales between incompatible units were also considered, but leaving everything in natural units proved to be sufficient for our purposes.

We used an iterative process to determine best clustering parameters for each region to cleanly separate all the substructure that was present on the visual examination of the data, and to recover the full extent of previously known populations. Due to a variety of different options that can be incorporated, such as scaling of the data, the range of the data, the parameters of HDBSCAN, there is never a single ``definitive'' approach to clustering: multiple implementation can be equally effective in tracing the underlying density distribution - however there are differences in the degree of spurious structures that may be introduced in the background, the degree to which various substructures get separated or merged together, or sensitivity to the stars found on the outer periphery of various clusters \citep{hunt2021} Out of all the perturbations of clustering that we have examined most optimal clustering in Perseus was achieved using the `leaf' method, as `eom' was most susceptible to merging of various known clusters (e.g, Autochthe and NGC 1333 and/or Alcaeus) into a single population due to their similarity in phase space which was not considered ideal in our analysis. We have also examined the effect of cluster and sample sizes: setting the sample size too low produces too many spurious groups, setting it too high results in missing real clusters with fewer stars (e.g. NGC 1333). Setting the cluster size allows to reject the bulk of the spurious groups introduced by a low sample size, but setting it too high leads to excessive merging of substructures. Following various trials, we set the minimum sample size to 10 stars, with the minimum cluster size of 25 stars.

\subsection{Ages}

\begin{figure*}
\plotone{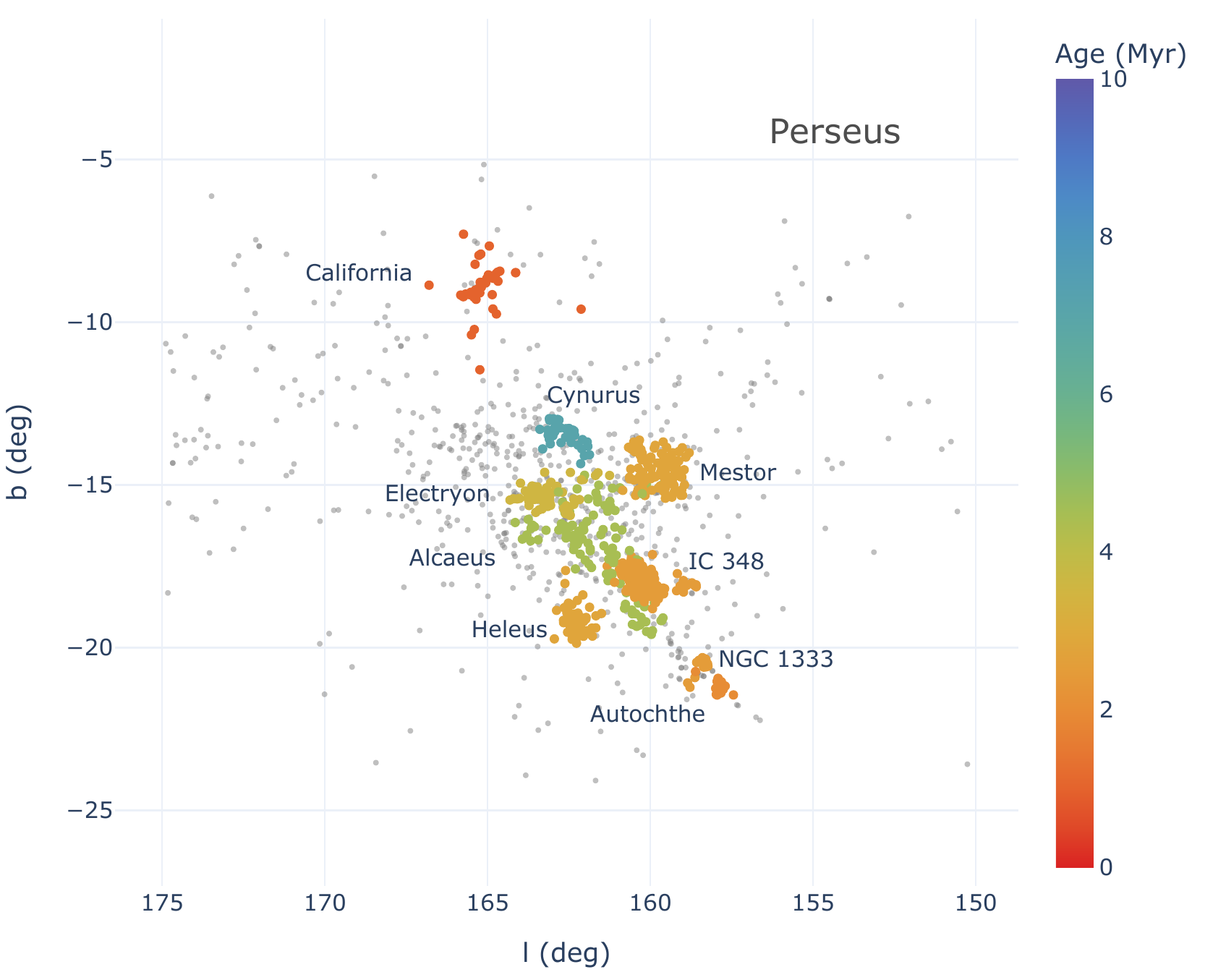}
\caption{Spatial distribution on the sky of 810 stars in the identified groups towards Per OB2, color-coded by their age. Grey dots correspond to 1171 PMS stars from \citet{mcbride2021} with PMS probability $>$90\% within the same volume of space.
\label{fig:sky}}
\end{figure*}

The clustering analysis described in Section~\ref{sec:method} has produced 43 groups in Perseus. Most of them are older moving groups unrelated to the more recent star forming history in the region, or they consist of chance alignment of random field stars in the neighboring phase space. Then, through examining the resulting HR diagram, we selected groups younger than $<$7.5 dex. This further reduces the sample to only 9 groups (Figure \ref{fig:sky}).

The remaining 34 groups are not of interest to the analysis of recent star forming history in Perseus, and thus they are not pursued further. However some of the known clusters and populations have been recovered within this phase space, such as COIN-Gaia 10 \citep{cantat-gaudin2019}, as well as Theia 534, 733, 933, 1539, 1394 \citep{kounkel2019}. Many others do not necessarily form stable groups between different HDBSCAN runs with different hyperparameters, and may be spurious.

We estimate the ages of each of the identified young groups through manually performing isochrone fitting of the Gaia photometry to the MIST isochrones \citep{choi2016} as shown in Figure \ref{fig:hr}. Note that different set of isochrones may have a systematically different calibration in ages, but relative ages of different populations with respect to each other are typically consistent.

In addition to deriving average ages through isochrone fitting, we also consider ages of the individual stars, which are estimated with Sagitta \citep{mcbride2021}, it interpolates across Gaia and 2MASS photometry of individual pre-main sequence (PMS) stars with known ages from the literature, and it constructs an empirical age sequence. Additionally, it has a classifier that determines a probability of how likely is a particular star to be considered PMS in presence of more evolved field stars. It is effective for classifying young low mass stars, but it would generally return low probabilities for bona fide young stars of higher mass (due to their overlap with the red giant branch), those that are older than a few dozen Myr (due to their increased proximity to the field binary sequence), or those with an age $\lesssim$0.5 Myr (due to their rarity in the original training sample). We use both the classification and ages to perform a more detailed investigation of California cloud in Section \ref{sec:cali}, as it shows significant age spread.

\subsection{Average properties}

For each of the 9 identified young groups, we measure medians of the phase space they inhabit from the stars associated with them, in $l$, $b$, $\mu_l$, $\mu_b$, and $\pi$. Finally, although radial velocities (RVs) are not available for every star, there are generally several stars in each group from which an average can be computed.

We convert the medians of the phase space for each group to heliocentric rectangular X, Y, Z positions, combined with U, V, W velocities, with Z=0 corresponding to the Galactic plane, X increasing towards the Galactic center, and Y increasing in the direction of the Galactic rotation. We subtract the mean velocities from their UVW components to remove the orbital motion and to analyze relative velocities for a better interpretation of their star formation histories.

In Perseus, 7 of the 9 groups match up with those reported on by \citet{pavlidou2021}. Another coincides with the California cloud/LK H$\alpha$ 101. The last group is located somewhat outside of the area on the sky considered by \citet{pavlidou2021} in their analysis. Following their naming convention, we refer to it as Cynurus. It is the oldest of all 9 groups, with the age of 7 Myr. 

\subsection{Completeness \& Contamination}

\begin{figure}
\plotone{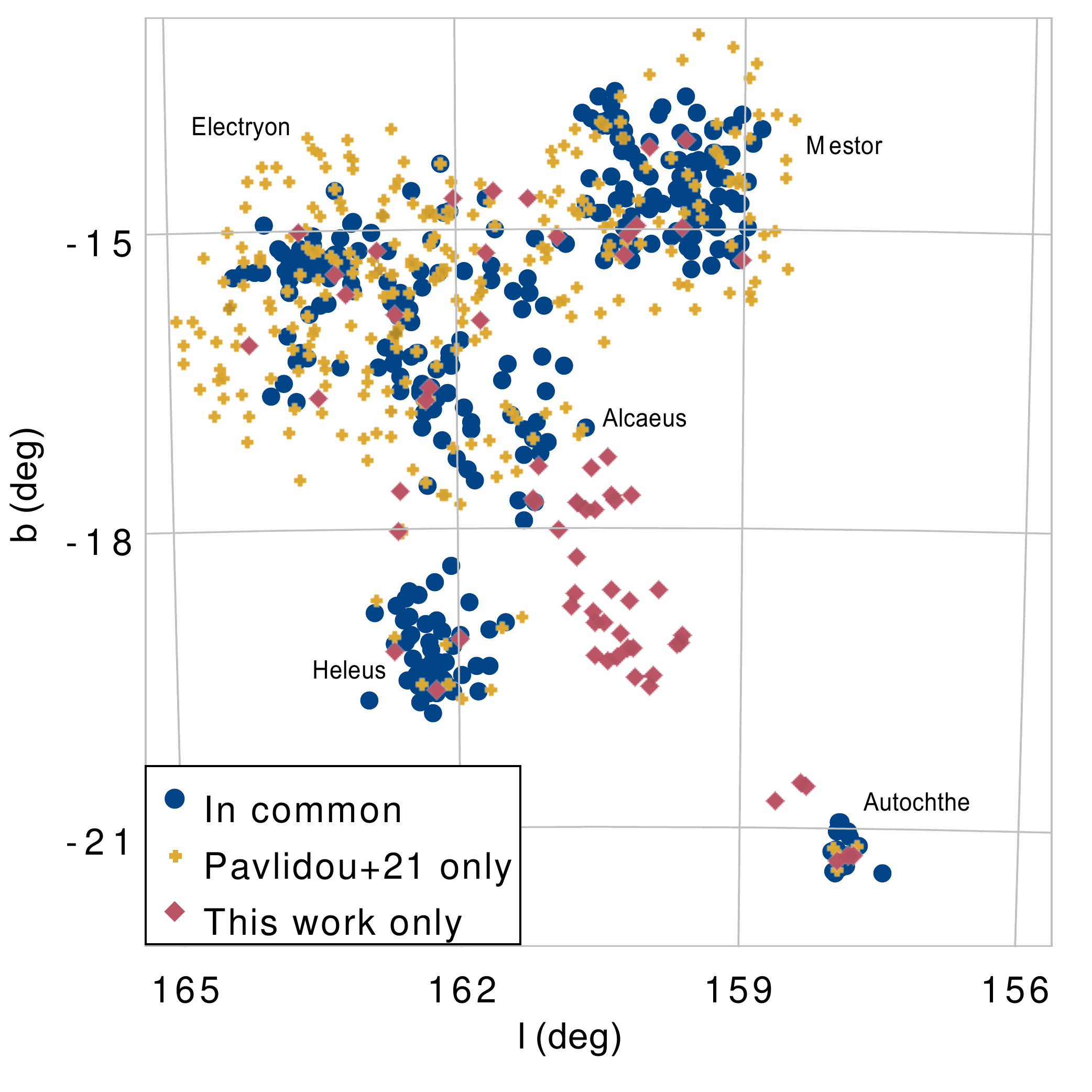}
\caption{A comparison of spatial distribution candidate members for five groups presented in \citet{pavlidou2021} (down to $G<$18 mag) and the corresponding groups in this work.
\label{fig:compare}}
\end{figure}

The 810 identified candidate members of these 9 young groups are listed in Table \ref{tab:individ}. The average properties of the groups are listed in Table \ref{tab:ave}. Comparing these properties are in a good agreement with \citet{pavlidou2021} for the groups that are in common between our samples. Their sample tends to have a larger number of stars in all the groups: the bulk of the difference can be explained by the limiting magnitude between studies. \citet{pavlidou2021} go down to $G<20$ mag, whereas we limit the clustered sample to only $G<18$ mag stars, as the photometry for the fainter stars tends to be less reliable, and the uncertainty in the astrometric parameters becomes large. Our catalog may also miss stars from \citet{pavlidou2021}  with $G$ between 17 and 18 mag found at large radii from the cluster center, particularly in the case of Electryon and Mestor. On the other hand, \citet{pavlidou2021} does not recover the full extend of morphology in Alcaeus that we do (Figure \ref{fig:compare}).

There is not a single comprehensive catalog of members of Per OB2 that does not suffer from some degree of incompleteness, as such it is difficult to infer the total number of sources that are missing from our analysis, but some estimates can be made. \citet{mcbride2021} have performed a survey of PMS stars across the entire sky selected on the basis of their photometry using Sagitta, however this sample lacks high mass stars. Performing a selection from their catalog with similar criteria described in Section \ref{sec:data}, and limiting the sources to PMS probability of $>$90\%, there appear to be 1171 young stars in the volume of space covered by our analysis. A number of these sources are distributed; performing a rough selection within an outline of where we recover clustered sources results in $\sim$1050 stars. Applying the classifier from Sagitta on the clustered sample presented here results in a sample of 529 stars, excluding the bulk of stars with [BP]-[RP]$<$2. As such, sources recovered through clustering may be missing as much as half of the likely members in the initial sample in Section \ref{sec:data}. The region of greatest incompleteness is near the top of Per OB2, in vicinity of Cynurus and Electryon, where the population is most dispersed. Given that the clustering recovers primarily regions of highest overdensity, sources along the more diffuse halo may be missed (Figure \ref{fig:sky}).

\begin{figure*}
		\gridline{
		     \fig{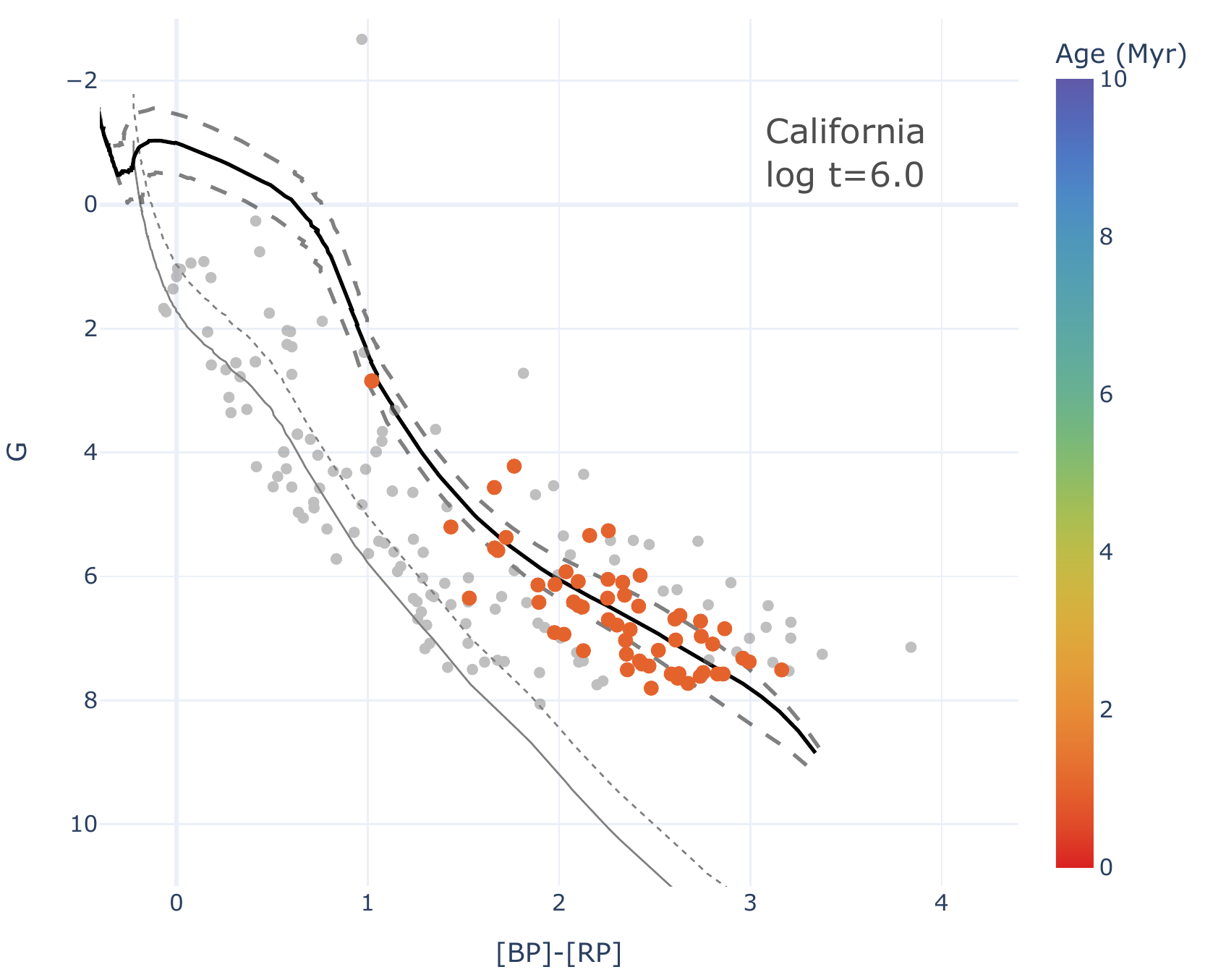}{0.33\textwidth}{}
             \fig{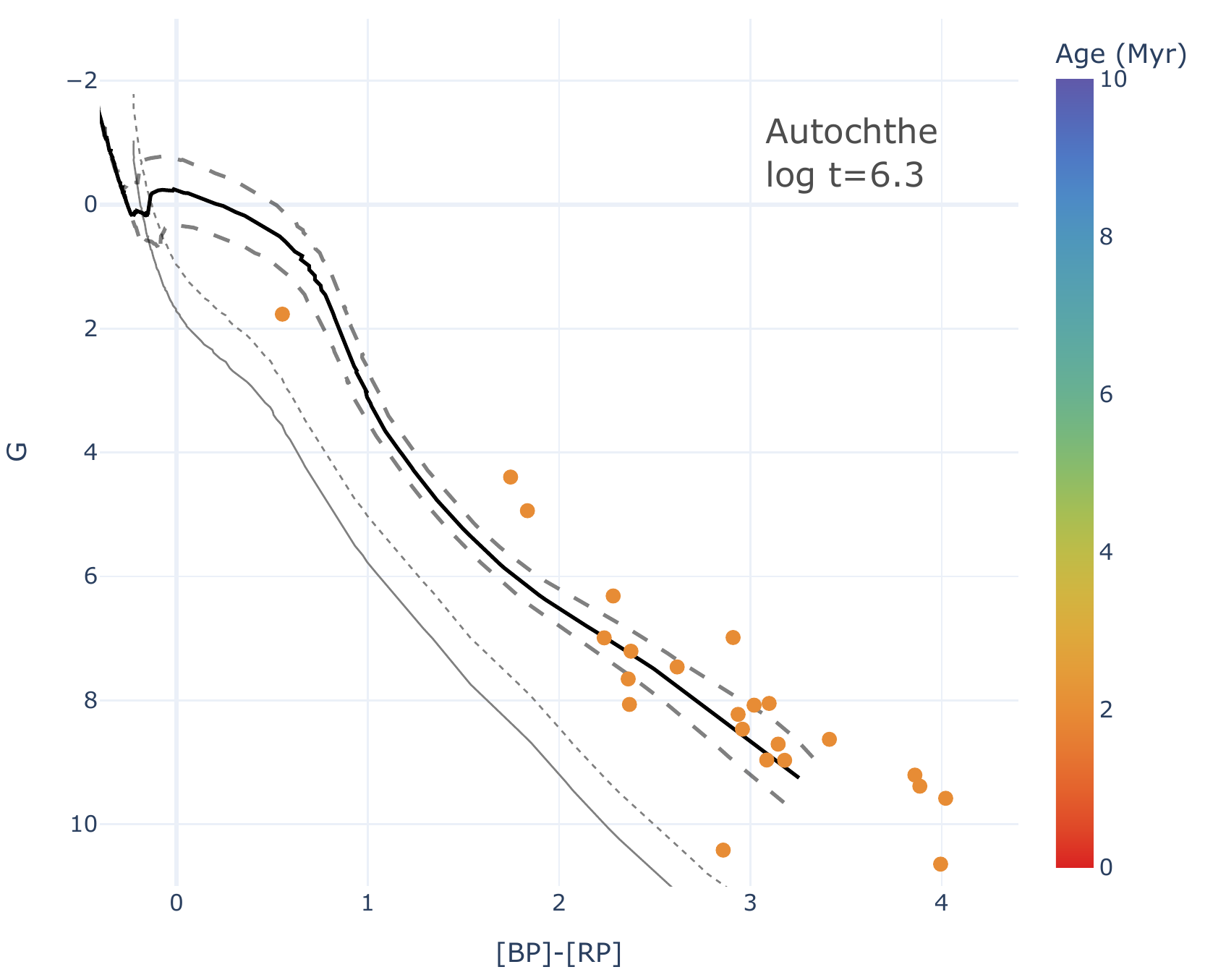}{0.33\textwidth}{}
             \fig{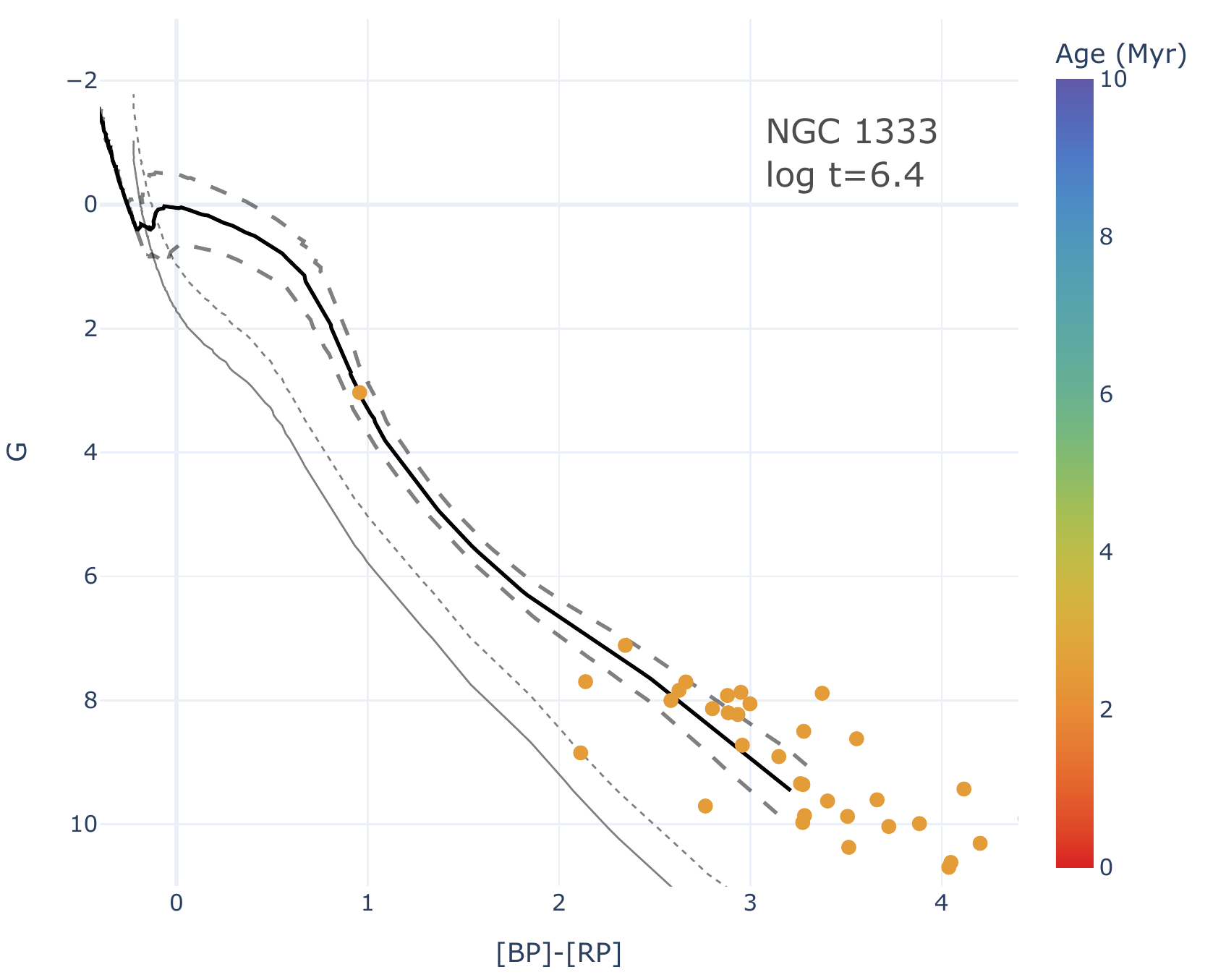}{0.33\textwidth}{}
        }\vspace{-1cm}
        \gridline{
		     \fig{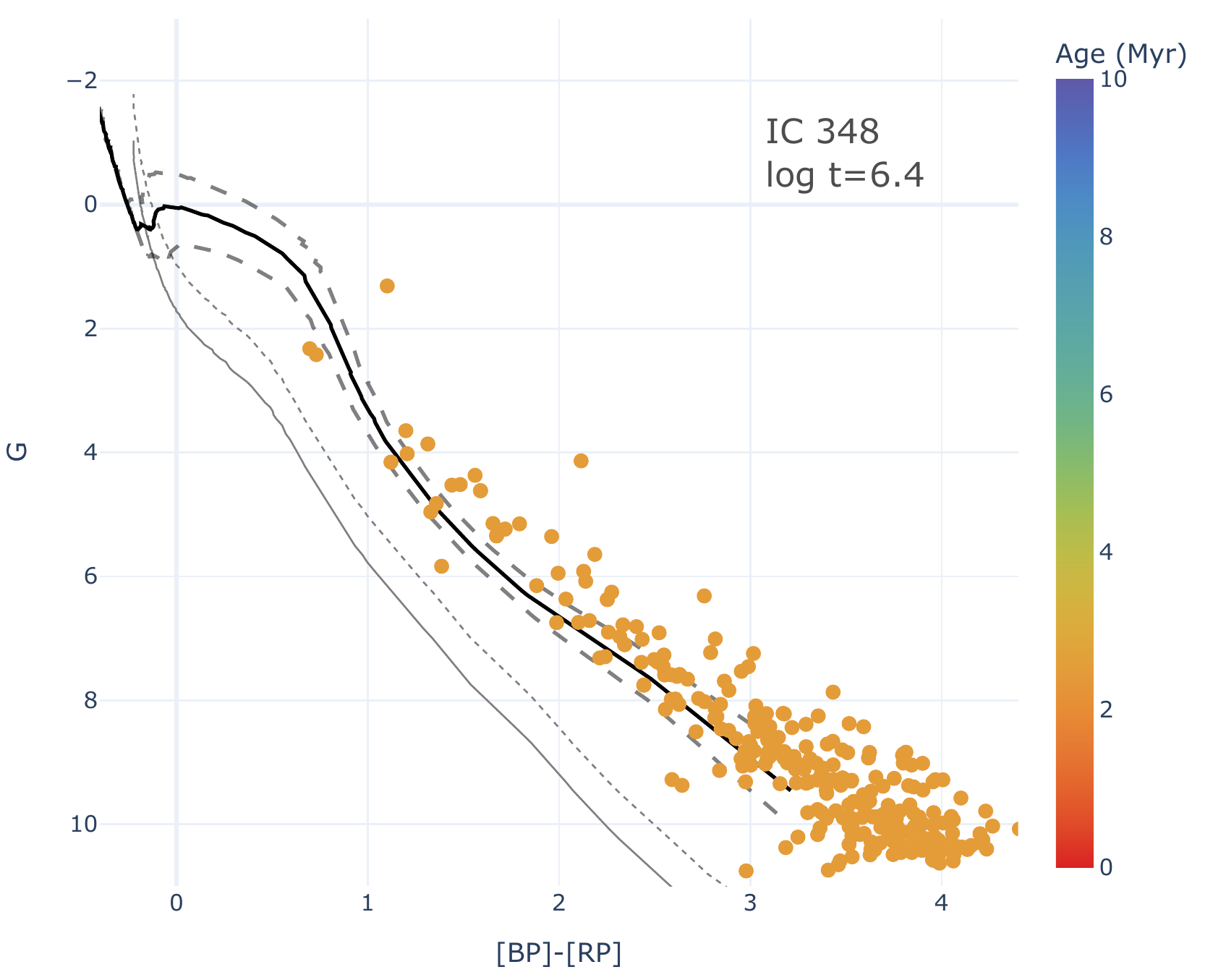}{0.33\textwidth}{}
             \fig{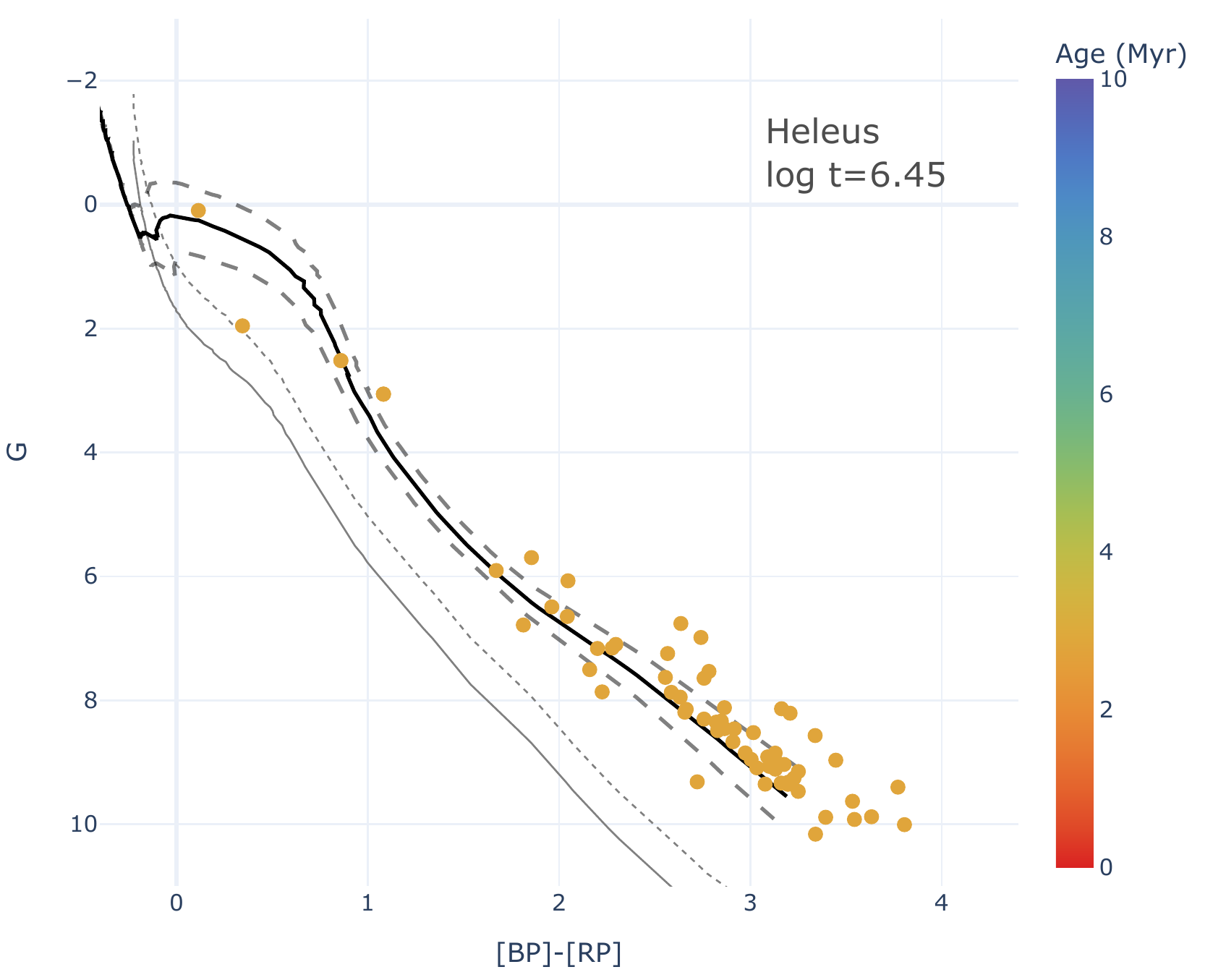}{0.33\textwidth}{}
             \fig{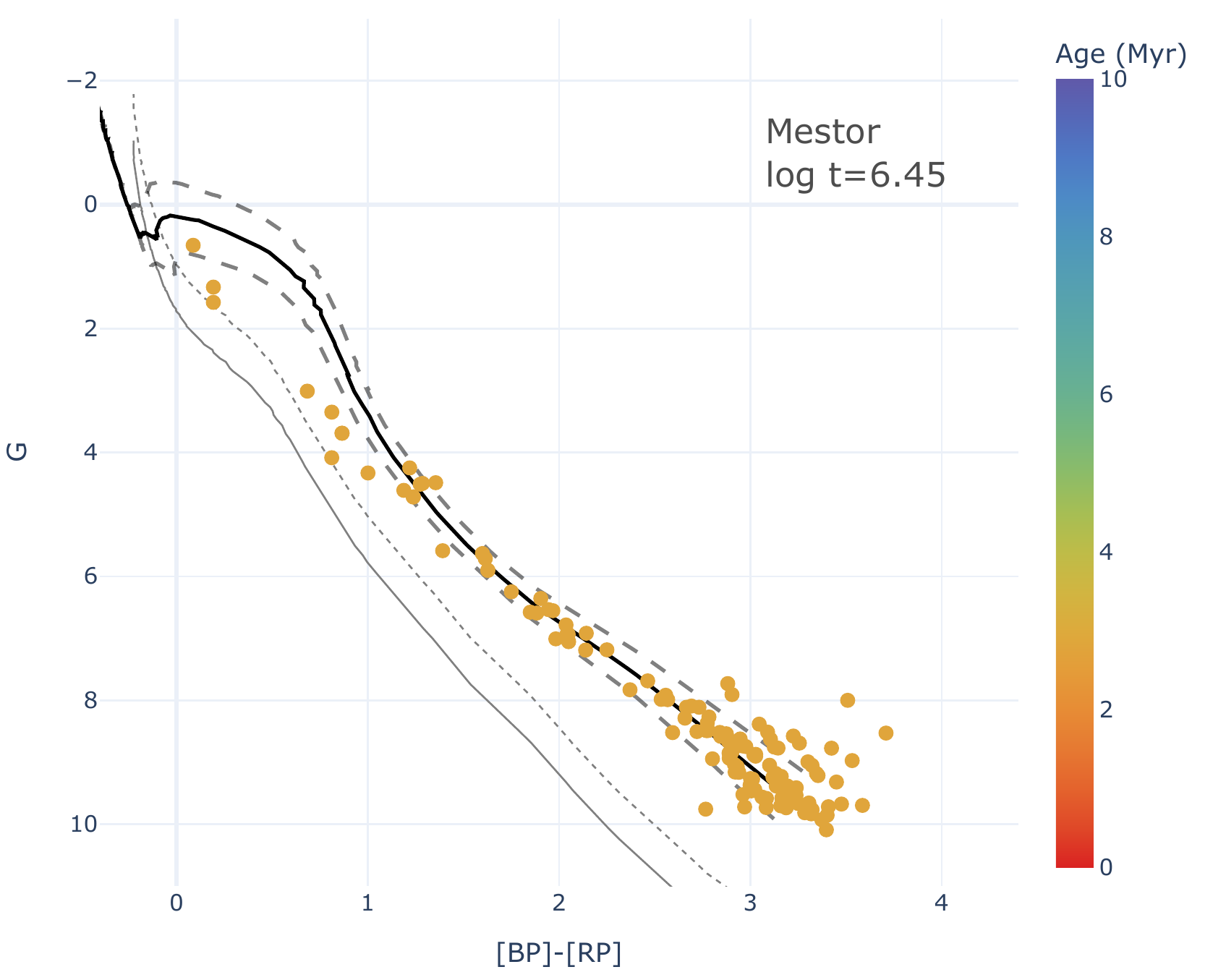}{0.33\textwidth}{}
        }\vspace{-1cm}
        \gridline{
		     \fig{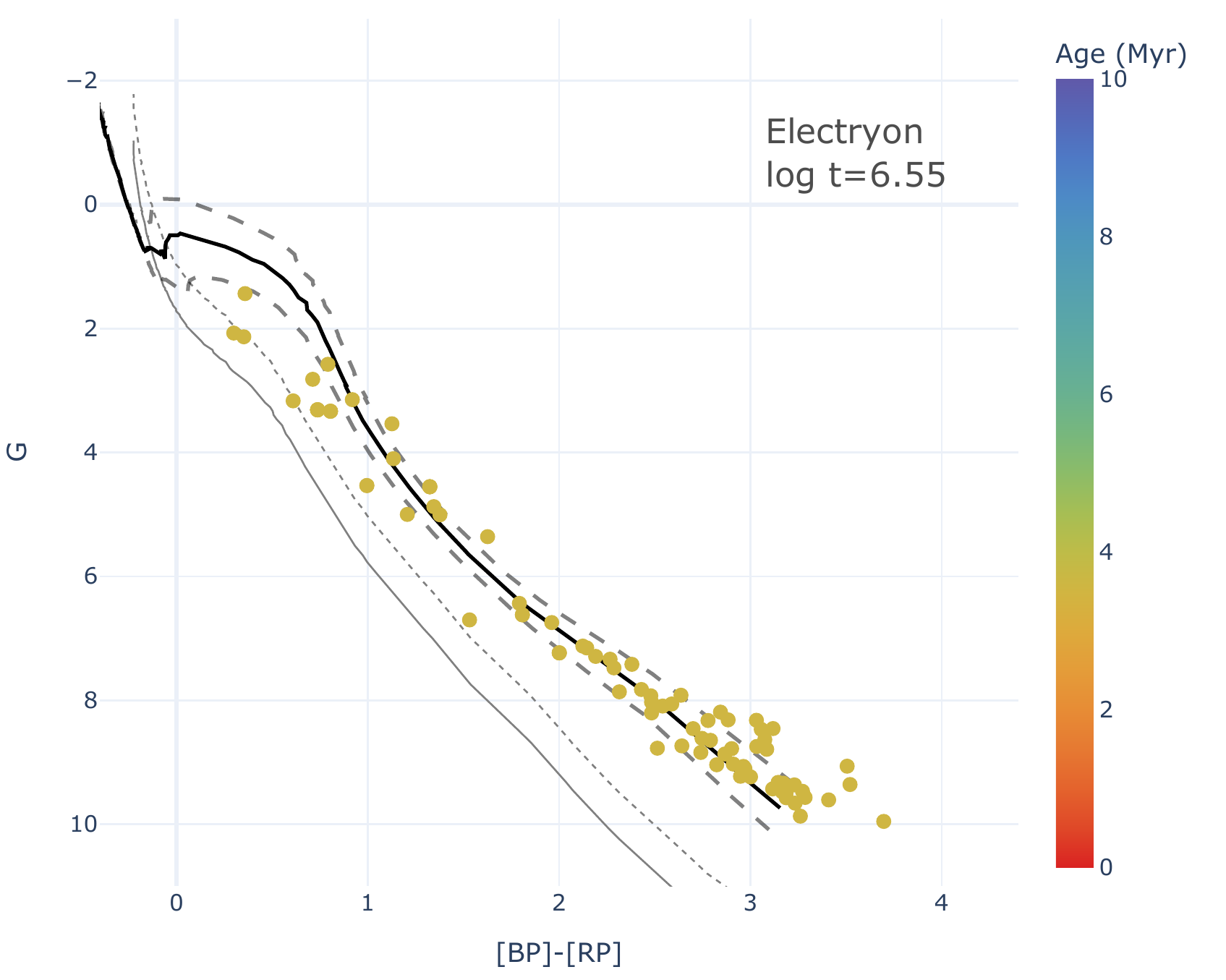}{0.33\textwidth}{}
             \fig{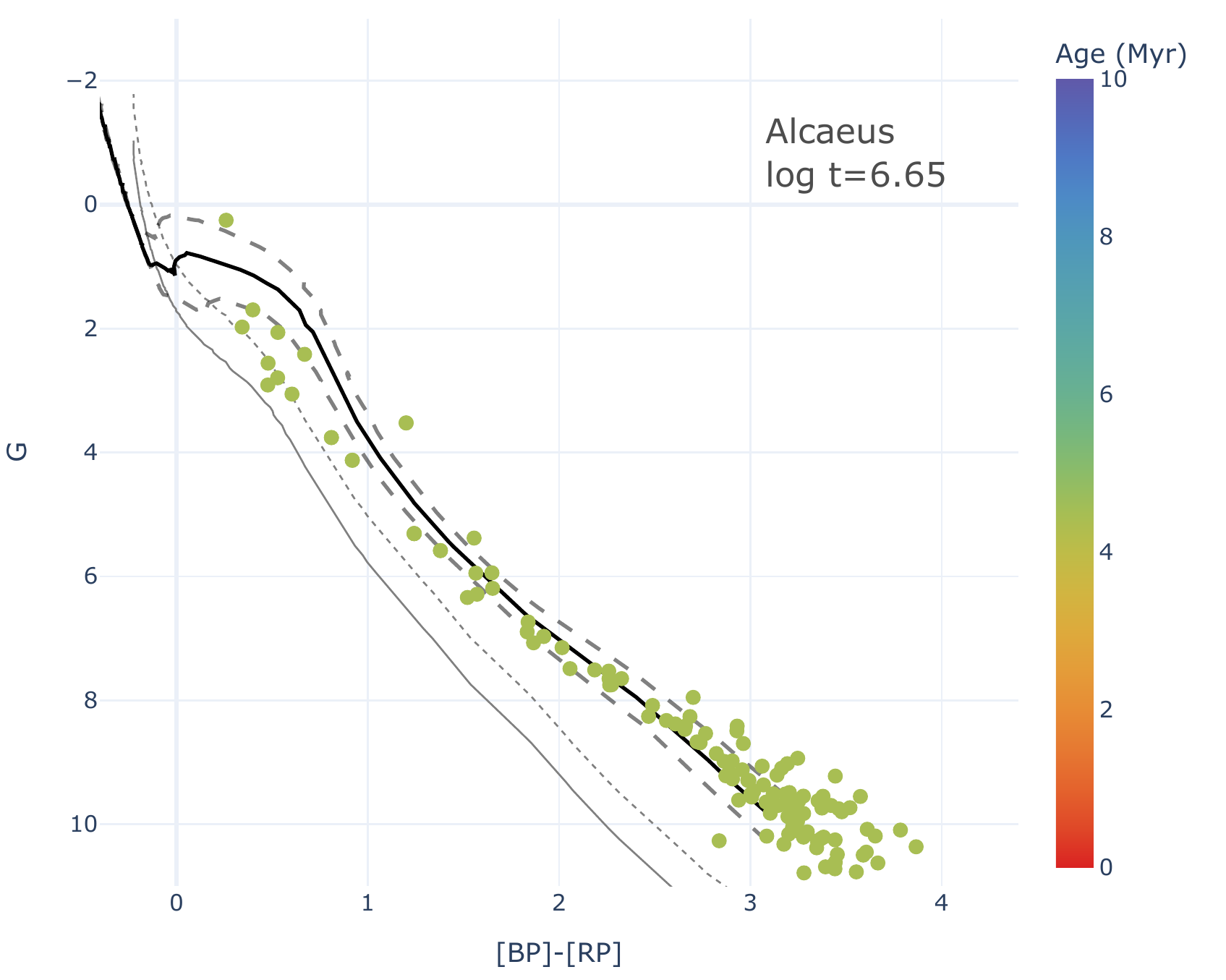}{0.33\textwidth}{}
             \fig{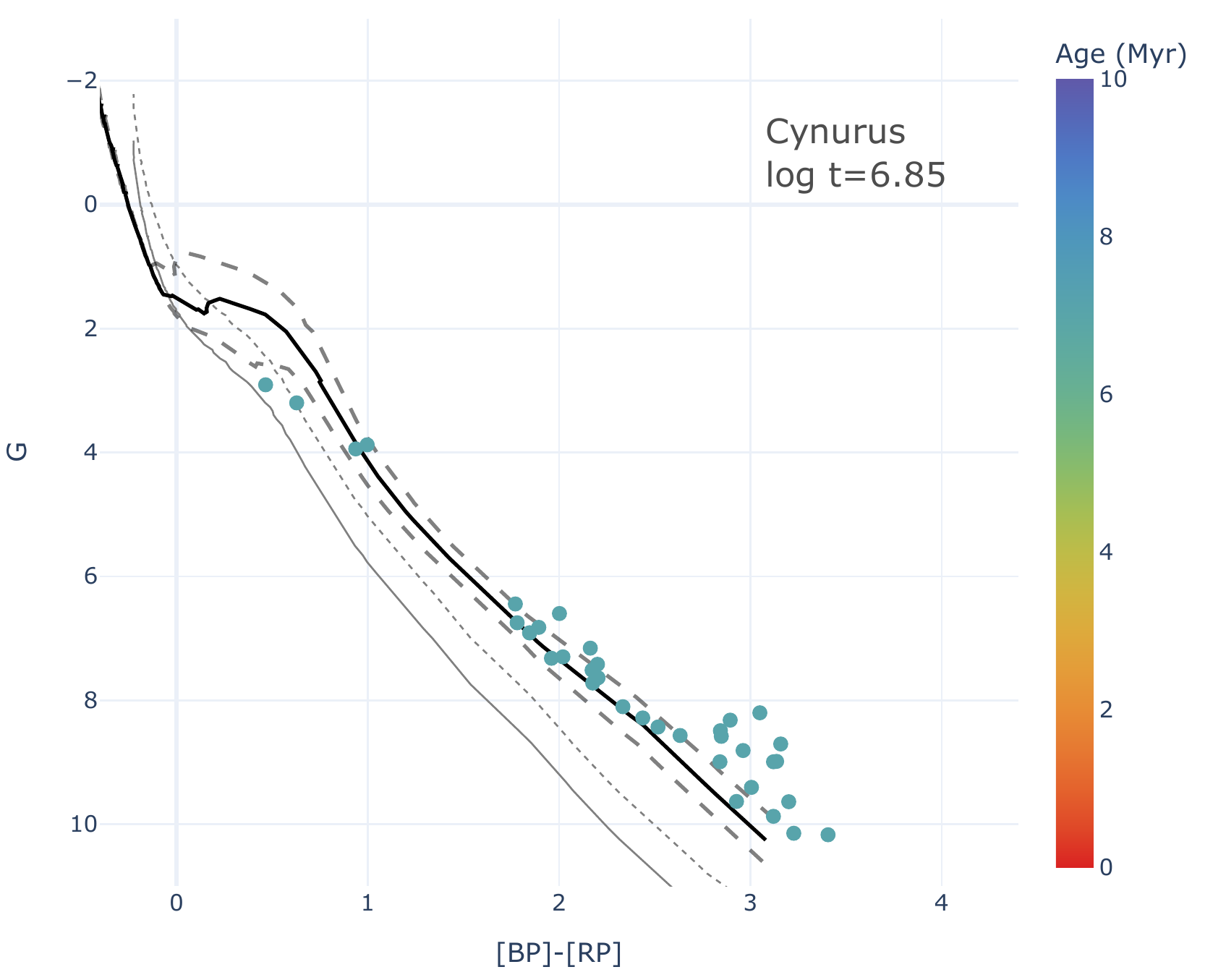}{0.33\textwidth}{}
        }\vspace{-1cm}
\caption{HR diagram for all the groups in the sample. The solid black line shows the best-fitted MIST isochrone \citep{choi2016}, from Table \ref{tab:ave}. The dashed lines show the isochrone 0.2 dex younger and 0.2 dex older. The thin grey line shows the main sequence, and the dotted thin line is the field binary sequence, included for the reference to estimate contamination. In California cloud, grey sources have PMS probability measured by Sagitta $<$90\% (Note that the classifier may also exclude the youngest stars with ages $\lesssim$0.5 Myr due to their rarity in the training set).
\label{fig:hr}}
\end{figure*}

Another consideration for the sample reported here is the fraction of the sample that may be contaminated by the field stars. Examining the HR diagram in Figure \ref{fig:hr} for every group shows that almost all of the sources within Per OB2 tend to follow relatively narrow cluster sequences, with only a minimal number of sources close to the main sequence, which would be the dominant source of contamination. As such, the selection appears to be relatively clean. On the other hand, contamination towards the California cloud is much more significant, with as many as 30--50\% of sources being near the main sequence. There may be an unrelated older moving group in a vicinity sharing similar kinematics, or a larger degree of opacity in the cloud combined with high opacity introduces contamination from unrelated field stars. As such, to refine the sample by applying Sagitta classifier, and selecting the sources with PMS probability $>$90\%, limiting the sample in this region to only 58 stars, and calculate the average properties only from them. The remaining groups do not have such a cut imposed, since they do not appear to have significant contamination, but such a cut would preferentially excludes higher mass stars.

%\begin{deluxetable*}{cccccccccc}
%\tablecaption{Candidate members of young groups in Perseus
%\label{tab:individ}}
%\tabletypesize{\scriptsize}
%\tablewidth{\linewidth}
%\tablehead{
%  \colhead{Gaia EDR3} &
%  \colhead{$\alpha$} &
%  \colhead{$\delta$} &
%  \colhead{$\pi$} &
%  \colhead{$\mu_\alpha$} &
%  \colhead{$\mu_\delta$} &
%  \colhead{RV} &
%  \colhead{RV} &
%  \colhead{Cluster} &
%  \colhead{PMS}\\
%  \colhead{source id} &
%  \colhead{(J2000)} &
%  \colhead{(J2000)} &
%  \colhead{(mas)} &
%  \colhead{(\masyr)} &
%  \colhead{(\masyr)} &
%  \colhead{(\kms)} &
%  \colhead{source} &
%  \colhead{} &
%  \colhead{prob.}
%  }
%\startdata
%  120922678211135616 & 52.176848 & 30.498055 & 3.464$\pm$0.051 & %6.843$\pm$0.065 & -9.431$\pm$0.046 & 12.315$\pm$0.094 & APOGEE & %NGC 1333 & 0.93\\
%  219994379592952320 & 57.589364 & 36.110374 & 2.665$\pm$0.089 & %3.29$\pm$0.13 & -4.492$\pm$0.087 &  &  & Mestor & 0.92\\
%\enddata
%\tablenotetext{}{Only a portion shown here. Full table is available %in electronic form.}
%\end{deluxetable*}

\begin{deluxetable}{ccl}
\tablecaption{Candidate members of young groups in Perseus
\label{tab:individ}}
\tabletypesize{\scriptsize}
\tablewidth{\linewidth}
\tablehead{
  \colhead{Column} &
  \colhead{Unit} &
  \colhead{Description}
  }
\startdata
Source id & & Gaia EDR3 source ID \\
$\alpha$ & deg & Right ascention (J2000) \\
$\delta$ & deg & Declination (J2000) \\
$\pi$ & mas & Gaia EDR3 parallax \\
$\sigma_\pi$ & mas & Uncertainty in parallax \\
$\mu_\alpha$ & \masyr & Heliocentric proper motion in $\alpha$ \\
$\sigma_{\mu_\alpha}$ & \masyr & Uncertainty in $\mu_\alpha$ \\
$\mu_\delta$ & \masyr & Heliocentric proper motion in $\delta$ \\
$\sigma_{\mu_\delta}$ & \masyr &  Uncertainty in $\mu_\delta$ \\
RV & \kms& Heliocentric radial velocity \\
$\sigma_{\rm RV}$ & \kms & Uncertainty in RV \\
RV source & & Source of RV data \\
$l$ & deg & Galactic longitude \\
$b$ & deg & Galactic latitude \\
$v_{lsr,l}$ & \kms & LSR corrected tangential velocity in $l$ \\
$v_{lsr,b}$ & \kms & LSR corrected tangential velocity in $b$ \\
RV$_{lsr}$ & \kms & LSR corrected radial velocity \\
G & mag & G band magnitude \\
BP & mag & BP band magnitude \\
RP & mag & RP band magnitude \\
PMS & & Probability of a star being PMS \\
Group & & Group membership 
\enddata
\tablenotetext{}{Full table is available in electronic form.}
\end{deluxetable}

\begin{deluxetable}{ccccccccccccccccc}
\tablecaption{Identified groups
\label{tab:ave}}
\tabletypesize{\scriptsize}
\tablewidth{\linewidth}
%\colnumbers
\rotate
\tablehead{
  \colhead{Name} &
  \colhead{log $t$} &
  \colhead{$l$} &
  \colhead{$b$} &
  \colhead{Dist.} &
  \colhead{$\mu_{l,lsr}$} &
  \colhead{$\mu_{b,lsr}$} &
  \colhead{RV$_{lsr}$} &
    \colhead{X} &
    \colhead{Y} &
    \colhead{Z} &
    \colhead{U$_{lsr}$} &
    \colhead{V$_{lsr}$} &
    \colhead{W$_{lsr}$} &
    \colhead{N} &
    \colhead{NRV} &
    \colhead{Source} \\
  \colhead{} &
  \colhead{[yr]} &
  \colhead{(deg.)} &
  \colhead{(deg.)} &
  \colhead{(pc)} &
  \colhead{(\masyr)} &
  \colhead{(\masyr)} &
  \colhead{(\kms)} &
    \colhead{(pc)} &
    \colhead{(pc)} &
    \colhead{(pc)} &
    \colhead{(\kms)} &
    \colhead{(\kms)} &
    \colhead{(\kms)} &
    \colhead{} &
    \colhead{} & 
    \colhead{}
  }
\startdata
   California & 6.0 & 165.3$\pm$0.6 & -9.1$\pm$0.6 & 529$\pm$37 & -0.44$\pm$0.57 & 0.71$\pm$0.47 & -1.64$\pm$1.8 & -505 & 133 & -83 & 1.58 & 0.72 & 2.01 & 56 [186]\tablenotemark{$^a$} & 17 [70]\tablenotemark{$^a$} & This work\\
   \hline
  Autochthe & 6.3 & 157.9$\pm$0.2 & -21.2$\pm$0.2 & 295$\pm$10 & 0.35$\pm$0.26 & 1.67$\pm$0.24 & 5.67$\pm$4.94 & -255 & 103 & -107 & -5.87 & 1.86 & 0.12 & 25 & 7 & This work\\
   &  & 157.9 $\pm$ 0.4 & -21.3 $\pm$ 0.4 & 298$\pm$24 & 0.26 $\pm$ 1.1 & 1.07 $\pm$ 1.1 &  &  &  &  &  &  &  & 27 (19)\tablenotemark{$^b$} &  & \citep{pavlidou2021}\\
   \hline
  NGC 1333 & 6.4 & 158.3$\pm$0.1 & -20.5$\pm$0.2 & 294$\pm$6 & 0.26$\pm$0.56 & -0.23$\pm$0.33 & 6.58$\pm$4.43 & -256 & 102 & -103 & -5.76 & 1.90 & -2.59 & 34 & 27 & This work\\
  \hline
  IC 348 & 6.4 & 160.5$\pm$0.4 & -17.8$\pm$0.3 & 315$\pm$16 & -2.80$\pm$0.65 & 1.30$\pm$0.52 & 7.39$\pm$5.09 & -283 & 100 & -96 & -5.79 & 6.49 & -0.42 & 270 & 202 & This work\\
  \hline
  Heleus & 6.45 & 162.3$\pm$0.3 & -19.4$\pm$0.4 & 400$\pm$19 & -2.50$\pm$0.33 & 0.19$\pm$0.28 & 16.17$\pm$6.21 & -360 & 115 & -133 & -13.20 & 9.20 & -5.00 & 66 & 4 & This work\\
   &  & 162.3 $\pm$ 0.7 & -19.4 $\pm$ 0.7 & 413$\pm$31 & -2.11 $\pm$ 0.68 & 0.12 $\pm$ 0.68 &  &  &  &  &  &  &  & 85 (61)\tablenotemark{$^b$} &  & \citep{pavlidou2021}\\
   \hline
  Mestor & 6.45 & 159.6$\pm$0.5 & -14.6$\pm$0.5 & 390$\pm$15 & -2.92$\pm$0.28 & 1.84$\pm$0.19 & 11.67$\pm$6.25 & -354 & 132 & -98 & -9.51 & 9.31 & 0.35 & 124 & 18 & This work\\
   &  & 159.7 $\pm$ 1.2 & -14.6 $\pm$ 1.2 & 395$\pm$16 & -2.75 $\pm$ 0.73 & 1.65 $\pm$ 0.73 &  &  &  &  &  &  &  & 302 (121)\tablenotemark{$^b$} &  & \citep{pavlidou2021}\\
   \hline
  Electryon & 6.55 & 163.3$\pm$0.6 & -15.4$\pm$0.3 & 371$\pm$18 & -2.34$\pm$0.32 & 1.55$\pm$0.19 & 12.67$\pm$7.18 & -342 & 103 & -98 & -11.21 & 7.65 & -0.73 & 75 & 15 & This work\\
   &  & 163.3 $\pm$ 1.2 & -15.4 $\pm$ 1.2 & 370$\pm$15 & -2.2 $\pm$ 0.8 & 1.26 $\pm$ 0.8 &  &  &  &  &  &  &  & 329 (70)\tablenotemark{$^b$} &  & \citep{pavlidou2021}\\
   \hline
  Alcaeus & 6.65 & 161.6$\pm$1.1 & -16.7$\pm$1.2 & 285.4$\pm$10 & -0.11$\pm$0.36 & 0.66$\pm$0.29 & 7.34$\pm$6.11 & -259 & 86 & -82 & -6.87 & 2.44 & -1.25 & 124 & 29 & This work\\
   &  & 162.2 $\pm$ 1.2 & -16.6 $\pm$ 1.2 & 291$\pm$13 & 0.31 $\pm$ 0.89 & 0.48 $\pm$ 0.89 &  &  &  &  &  &  &  & 170 (76)\tablenotemark{$^b$} &  & \citep{pavlidou2021}\\
   \hline
  Cynurus & 6.85 & 162.7$\pm$0.4 & -13.4$\pm$0.3 & 358$\pm$15 & -2.36$\pm$0.19 & 1.92$\pm$0.15 & 9.97$\pm$6.87 & -333 & 104 & -83 & -8.79 & 6.95 & 0.86 & 36 & 7 & This work\\
  %name &  & l & b & d & pml & pmb &  &  &  &  &  &  &  & N &  & \citep{pavlidou2021}\\
\enddata
\tablenotetext{a}{Number of sources with PMS$>$90\% flag. Total number of clustered sources is given in brackets.}\vspace{-0.3cm}
\tablenotetext{b}{Number of sources in common with this work is given in parenthesis.}
\end{deluxetable}

\section{Results: 3d Structure and velocity}\label{sec:results}

In this section, we discuss the implications for the star-forming history in Perseus based on the systemic motions of the groups uncovered in Section~\ref{sec:method}. For enhanced visualization of the motions discussed, the reader is encouraged to utilize the interactive versions of Figure~\ref{fig:per3d}, which can be accessed via the electronic version of the Journal. 

\subsection{Per OB2}

\begin{figure*}
\epsscale{1.0}
\plotone{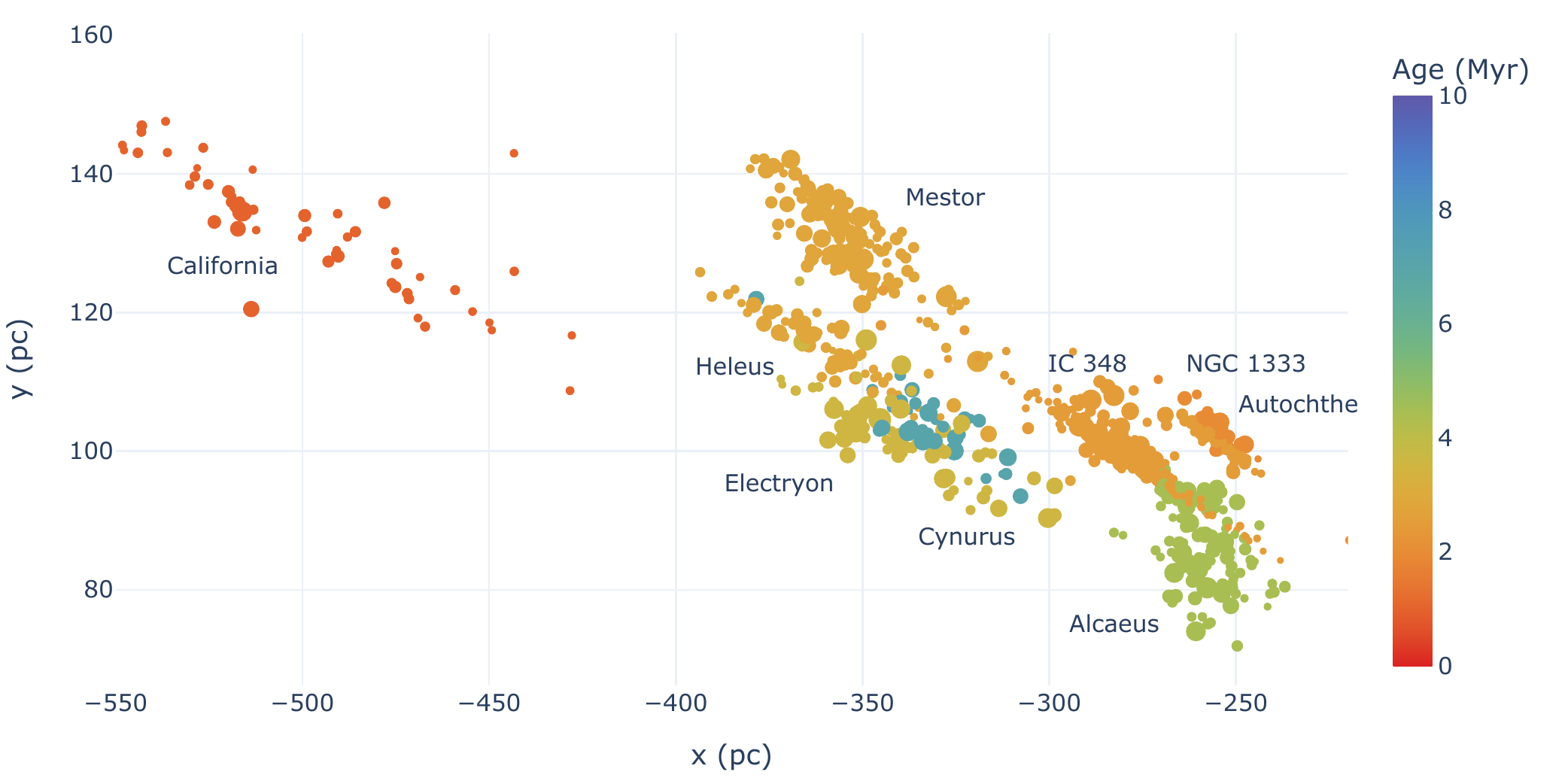}
\plotone{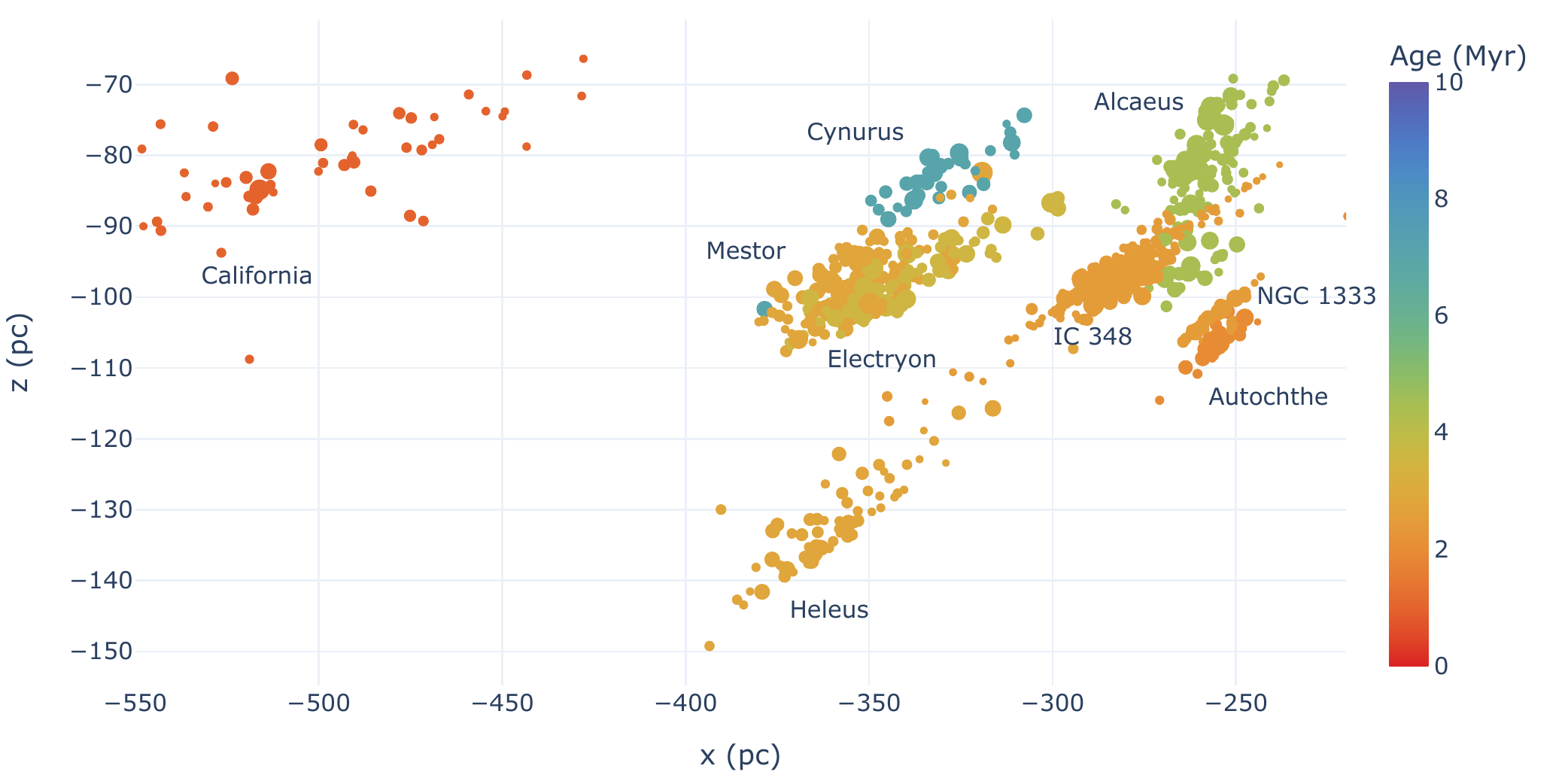}
\plotone{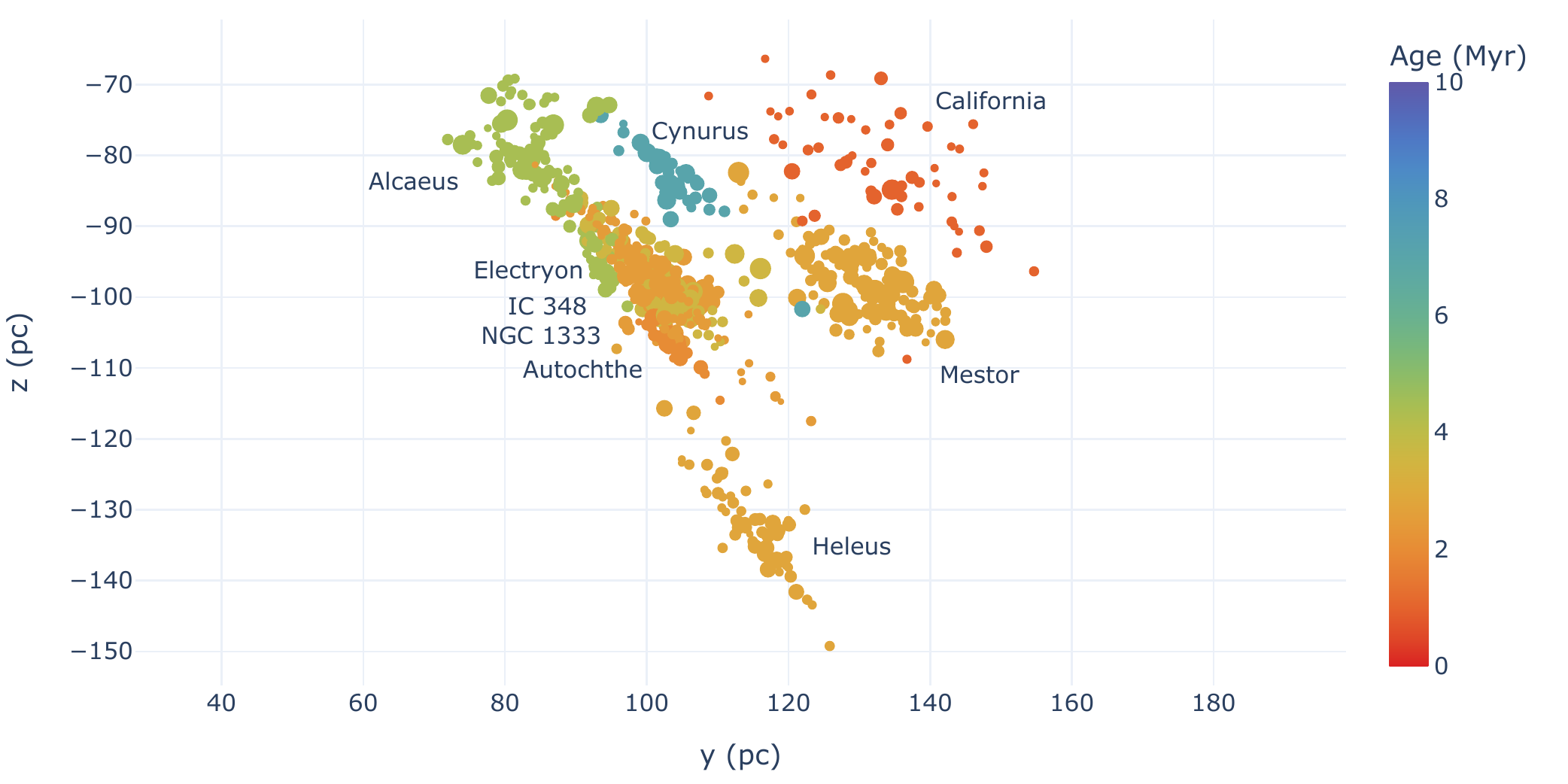}
\caption{Distribution of individual young stars towards Perseus along the 3-dimensional projection. Sources are color coded by the age of their parent group, similar to Figure \ref{fig:sky}. The symbol size is inversely correlated with the parallax uncertainty (ranging from $<$0.02 mas to 0.3 mas),  with larger symbols having a firmer distance measurement. 
\label{fig:per3dstar}}
\end{figure*}

\begin{figure*}
\begin{interactive}{js}{perseus.html}
\epsscale{1.0}
\gridline{\fig{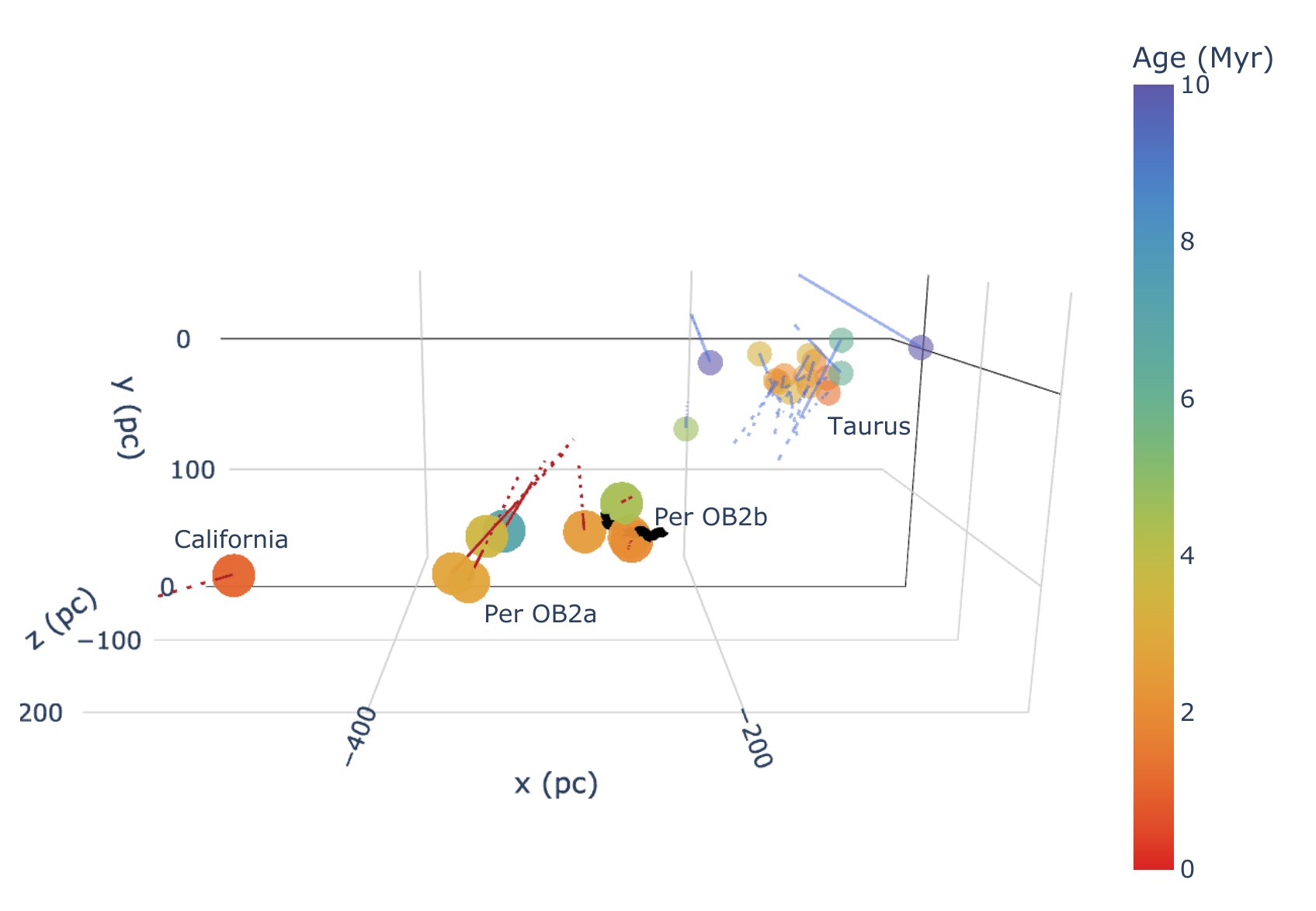}{0.75\textwidth}{}}\vspace{-2cm}
\gridline{\fig{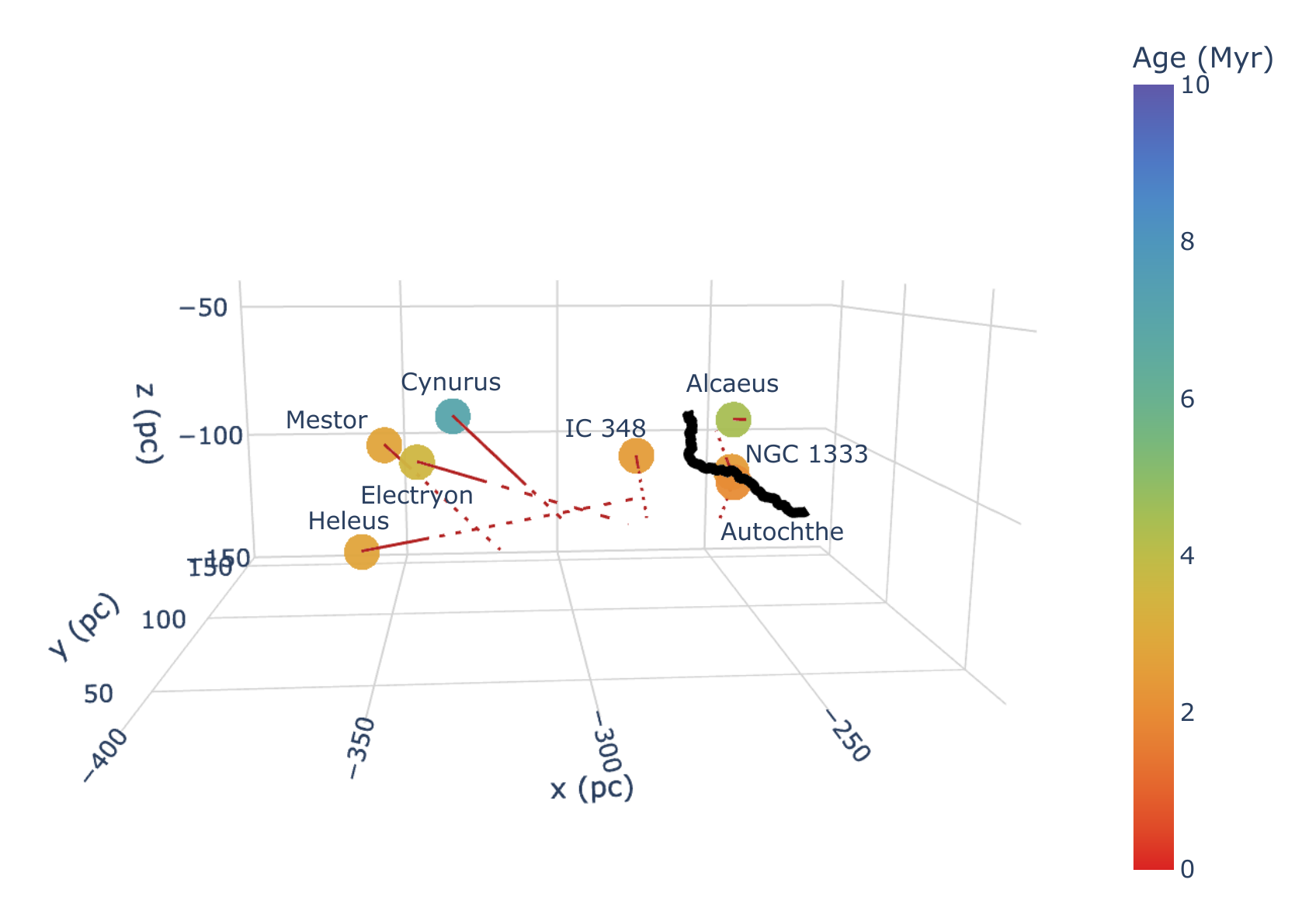}{0.75\textwidth}{}}
        \end{interactive}
\caption{Three-dimensional distribution of the identified groups towards Per OB2. The dots refer to the current position of all the groups, color-coded from their age. Lines emitting from them show the trace-back path over last 10 Myr, with solid portion of the line showing the path after the group has already formed, and dashed line - predating the bulk of star formation, when a population could have existed primarily as gas in molecular cloud, and they should be treated with caution. Top panel includes groups towards Taurus from \citet[smaller symbols with blue lines]{krolikowski2021}; the velocity reference frame is calculated relative to Per OB2b (NGC 1333, Autochthe, Alcaeus). Bottom panel is zoomed only on Perseus. Note that IC 348 is originating from the same point as Per OB2a, but it is moving towards Per OB2b. Black line shows the trace of the Perseus molecular cloud, from \citep[only shown for the current epoch]{zucker2021}; note the bend in the cloud relative Per OB2a. Interactive plot is available in the electronic version of the paper (temporarily \url{http://mkounkel.com/mw3d/perseus.html}), allowing rotation of the plot in 3d using dragging of the plot, zooming in using the scrolling wheel, and evolve the position of the groups back time using slider on the bottom of the plot. Dots become small once the timestep is older than the age of a group. The preset buttons in the interactive version allow changing the reference frame for the velocities (reference velocity is displayed in the corner), and to orient the plot in the plane of the sky. Clicking on the legend allows hiding certain traces.
\label{fig:per3d}}
\end{figure*}

Per OB2 has a complex morphology in the plane of the sky (Figure \ref{fig:sky}), with multiple groups distributed seemingly at random: they partially overlap, and most have no set pattern in their distribution. However, when viewed in 3d, it becomes apparent that the entire association does appear to have a shape of a filament, merely viewed nearly pole-on (Figure \ref{fig:per3dstar}).

However, as has been previously noted by \citet{kounkel2019} and by \citet{pavlidou2021}, this region has multiple distinct velocity components. Based on the classical name for the entire region, we refer to these two sets of groups as Per OB2a and Per OB2b. The former, Per OB2a consists of IC 348, Mestor, Heleus, Electryon, and Cynurus. The other group, Per OB2b, consists of NGC 1333, Autochthe, and Alcaeus.

Both Per OB2a and OB2b have their velocity vectors oriented in two near-opposite directions, with the relative angle between their trajectories of $\sim150^\circ$ along the X-Y plane in Figure \ref{fig:per3d}. However, this is not a signature of expansion. Rather, examining their velocities in 3d, their trajectories suggest an appearance of two populations that are passing each other by. As can be seen in the interactive plot, at the closest approach 5--6 ago Myr, the average center of Per OB2a and OB2b places them $\sim$70 pc apart, though, individual groups may approach closer, i.e, the center of IC 348 is only 20 pc away from the center of NGC 1333, and it will keep getting closer towards Per OB2b.

Within Per OB2, molecular gas is concentrated within the Perseus molecular cloud, and it is depleted elsewhere. NGC 1333 is still embedded in this cloud. A part of the Perseus molecular cloud is also found along the line of sight of IC 348, typically assumed to be associated with it. However, recently \citet{zucker2021}, have noted that the average parallax of the members of IC 348 versus the cloud distance along the line of sight towards this cluster are inconsistent with each other. IC 348 is located at a distance of $\sim$315--320 pc \citep[this work]{ortiz-leon2018,pavlidou2021}, whereas the cloud is found at the distance of 280--300 pc \citep{zucker2020,zucker2021}, and this difference in distances is larger than the typical uncertainty in the measurements. However, NGC 1333 and other groups that are a part of Per OB2b are also located at a distance similar to that of the molecular cloud $\sim$295--300 pc. This suggests that the molecular gas seen towards IC 348 and the cluster itself are not related and only appear in projection; rather the molecular gas is still present only in Per OB2b, with little of it remaining in Per OB2a.

When placed into a common reference frame, Per OB2a appears to be slightly expanding. As is can be seen in the interactive plot, the expansion age (i.e. the age at which all of the groups would have been at the closest proximity) exceeds 10 Myr, making it longer than the age of any of the groups withing it. Neither is this expansion entirely spherically uniform, while all of the groups within Per OB2a would have been closer in the past than they are now, they cannot be traced to a singular point from which all of them would originate at the same time. As this region lacks molecular gas, its total mass consists only of the stellar content, which amounts to only a few hundred \msun in total, which is not sufficient to keep the region gravitationally bound across the spatial scales covered by Per OB2a (even at the time of their closest approach), which results in its expansion. This expansion may have accelerated following the gas dispersal \citep{zamora-aviles2019}.

One of the outliers in Per OB2a is IC 348. While it certainly has a common origin with Per OB2a, it is located $\sim$50 pc away from its closest neighbor within it. Such a separation is comparable to the total size of the remaining Per OB2a groups combined, and it is due to IC 348 having the most discrepant velocity from the mean. Due to its closer proximity to Per OB2b, it appears to be most strongly influenced by its presence, and it is moving towards Per OB2b. As the latter still has sizable mass of $\sim1.3-1.6\times10^4$ \msun\ in form of the molecular gas \citep{ungerechts1987,zucker2021}, IC 348 may be infalling towards Per OB2b due to the gravitational potential \citep[a similar phenomenon has been previously observed in Orion][]{kounkel2018a,kounkel2020a}. Over 5 Myr, an object of this mass can accelerate an object located 20 pc away from rest to speeds of $\sim$1 \kms. Note that the mass of the individual clusters, including IC 348 itself, is over a magnitude smaller than that of the cloud. Combined with the initial expansion of Per OB2a, a presence of mass of the Perseus molecular cloud near one side would steepen the initial velocity gradient across Per OB2a. Due to the proximity of IC 348 to the Perseus molecular cloud, it would be the group most affected by the gravitational potential of a massive cloud near it. Other groups in Per OBa are $\sim$3 times further away from the cloud than IC 348 and thus they experience only a fraction of the force. On the other hand, groups within Per OB2b are already coincident with the cloud, having formed from it; the shape of the potential of the cloud they experience is significantly more affected by the precise density distribution within the cloud rather than its total mass.

Without detailed modeling of the initial conditions, it is difficult to say whether gravitational potential alone would be sufficient to fully reproduce the present day velocity structure, however, it is an important effect that should not be neglected. We note that while the Galactic potential is an important force, it would not dominate the potential of a massive molecular cloud on local scales of star forming regions, as such it would not prevent local infall from developing \citep{kounkel2021a}. We also consider other possible effects that may have shaped the kinematics in this region in Section \ref{sec:discussion}.

\subsection{California Cloud}\label{sec:cali}

A special mention should be given to California Cloud, which is not traditionally associated with Per OB2, as it is found at a larger distance, separated by $\sim$150 pc. It is on the trajectory towards Per OB2, and it may be able to reach it in the next $\sim$8--10 Myr. Due to such a large distance and an uncertainty in total mass, it is unclear if this is a signature of infall or a chance alignment.

\begin{figure*}
		\gridline{
             \fig{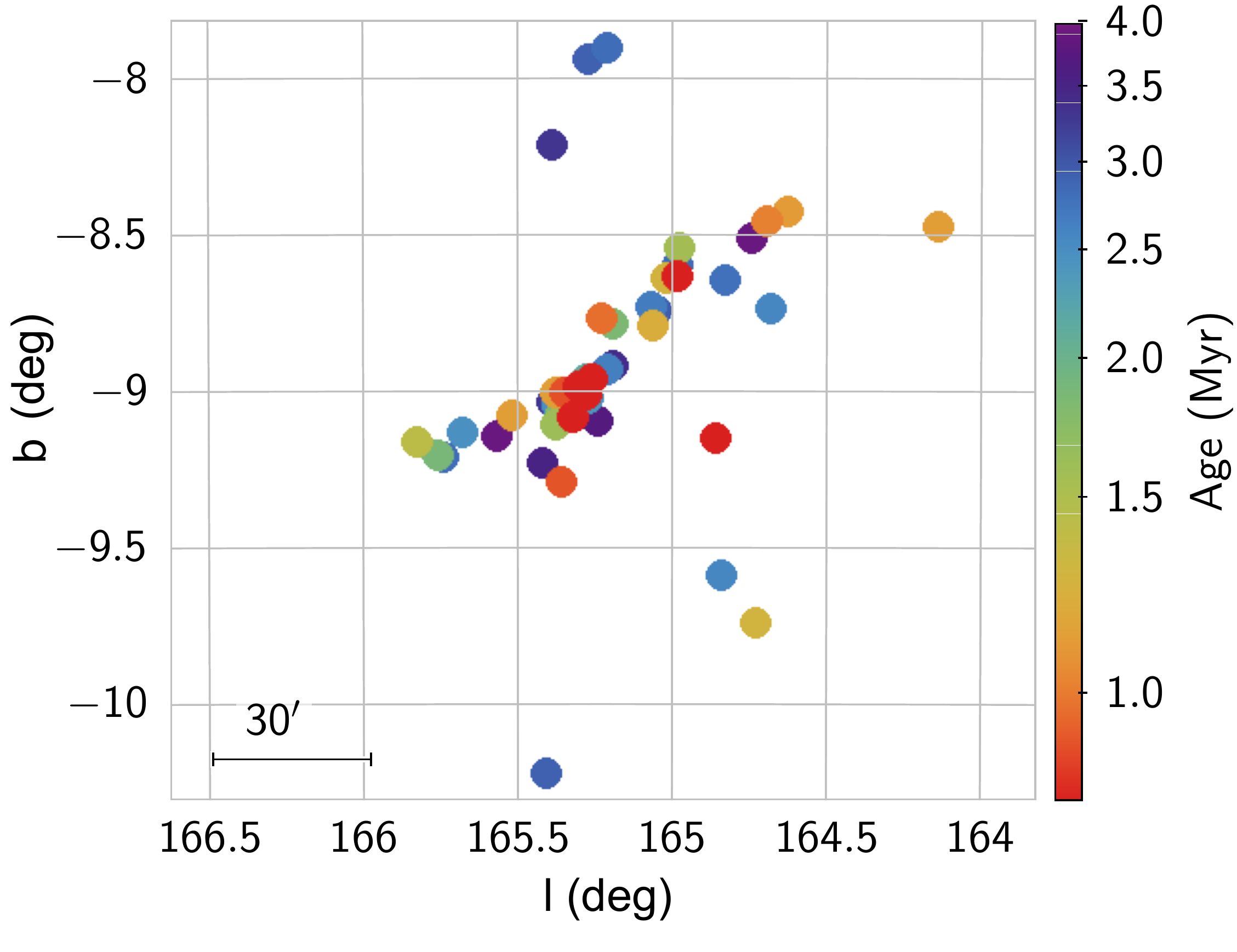}{0.5\textwidth}{}
             \fig{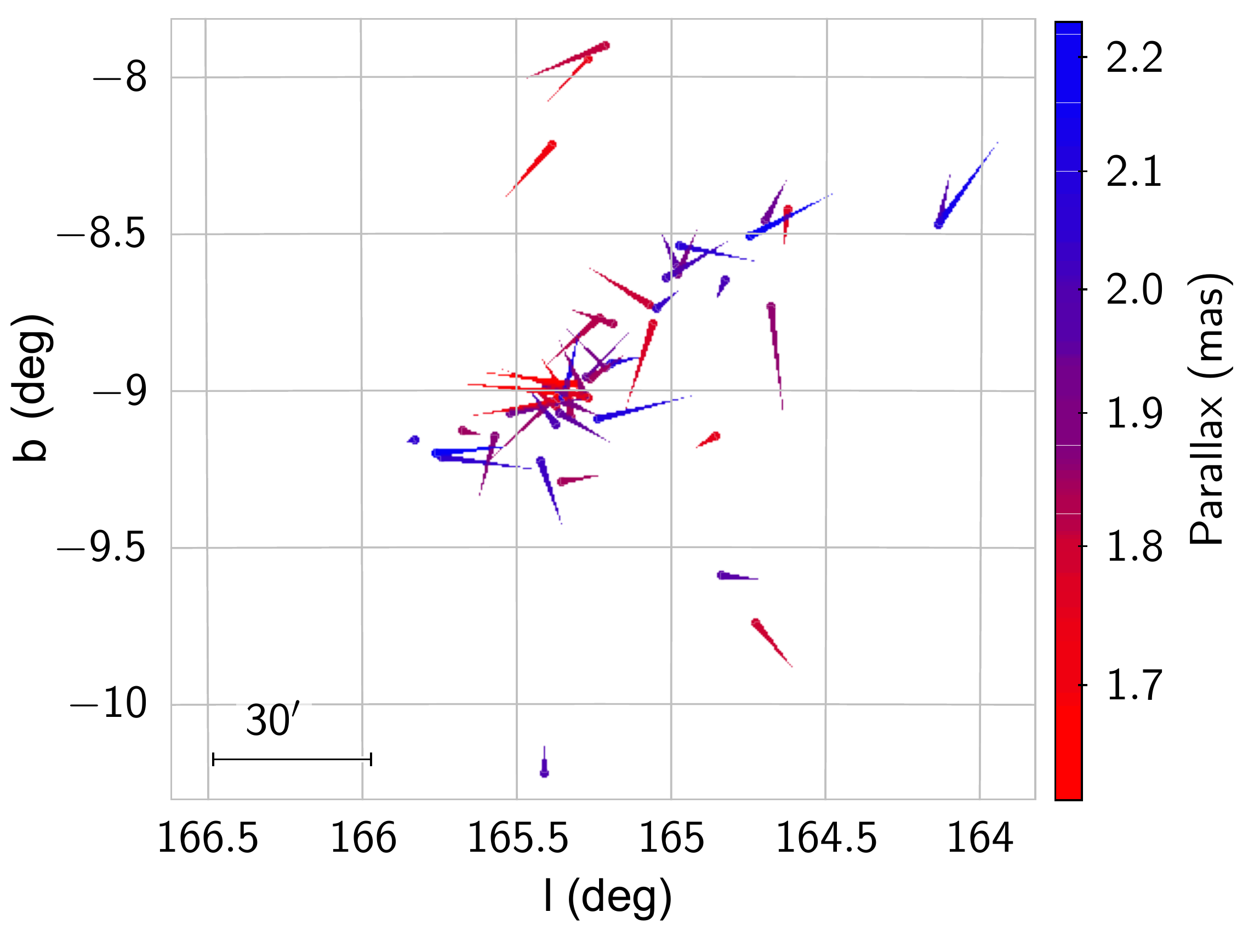}{0.5\textwidth}{}
        }\vspace{-1cm}
		\gridline{
             \fig{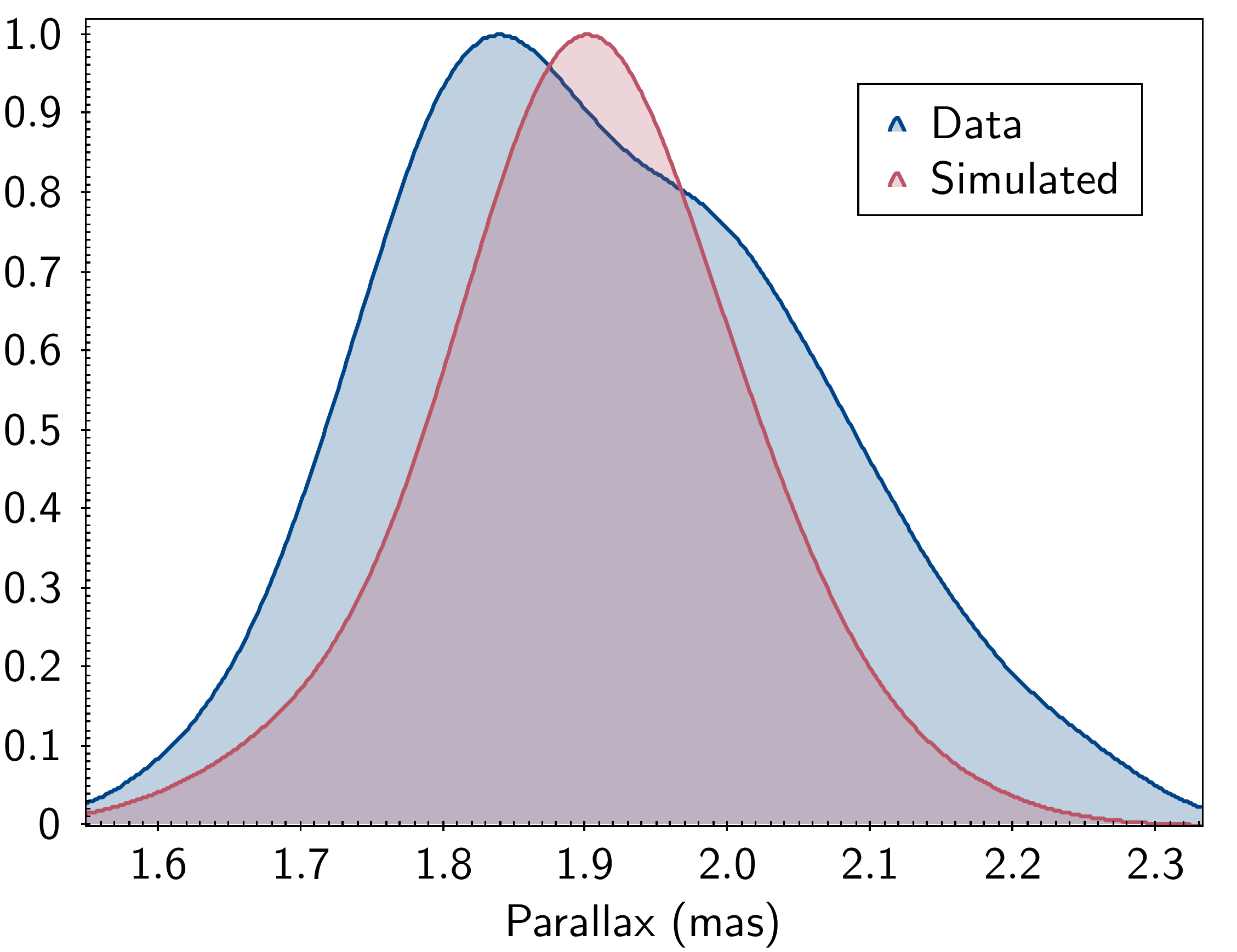}{0.5\textwidth}{}
             \fig{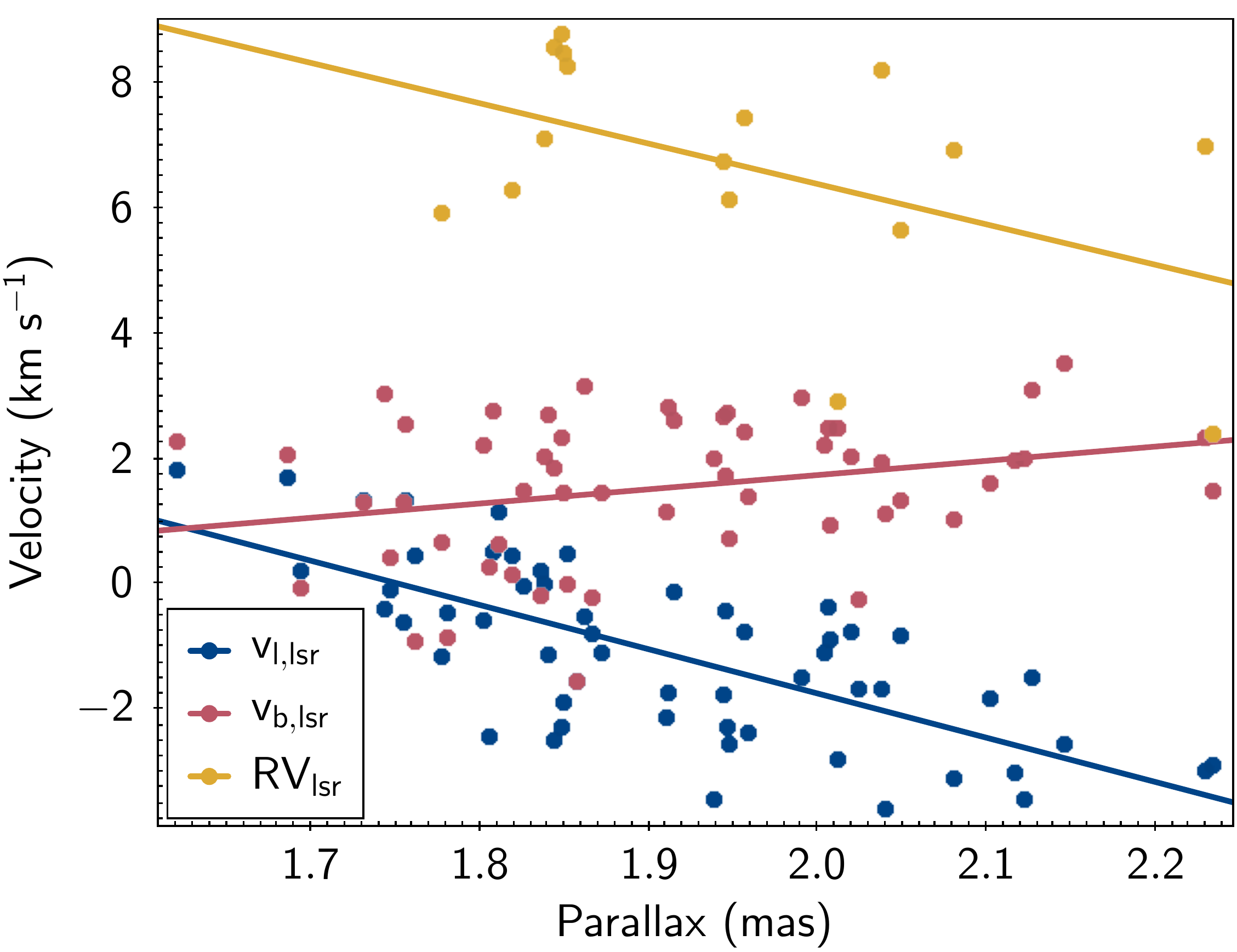}{0.5\textwidth}{}
        }
\caption{Top left: Distribution of candidate members identified towards California Cloud, color coded by the age estimated via Sagitta  \citep{mcbride2021}. Top right: proper motion of stars in the reference frame of California cloud, color coded by their parallax. Bottom left: Kernel density estimate of the overall distribution of parallaxes of stars within California cloud in blue. Red curve shows the simulated distribution of parallaxes for a compact population at a typical distance of California cloud, broadened by the parallax uncertainties of the stars. Bottom right: velocity gradients with distance of all three dimensions of motion.
\label{fig:cali}}
\end{figure*}

As can be seen in Figure \ref{fig:hr}, while we adopt a ``typical'' age for the California Cloud as 1 Myr, it has a significant spread in ages. Unlike other populations, even after limiting the sample only to the highest confidence PMS sources, it cannot be characterize well by a single isochrone. We can infer ages of individual stars using Sagitta \citep{mcbride2021}, their distribution is shown in Figure \ref{fig:cali} (top left).

We find an age gradient of stars towards California cloud of $\sim$4 Myr. The older stars tend to be found on the halo around the cloud. Youngest $\lesssim$1 Myr stars tend to be concentrated in a more compact hub at $l=165.4^\circ$, $b=-9^\circ$.

The identified candidate members span a wide distance distribution of $\sim$150 pc, from 450 pc to 600 pc It is comparable to the overall width of Per OB2 (combining both Per OB2a and Per OB2b), but it is much wider than what is expected of a single filamentary cloud or a single cluster. The uncertainties in parallax cannot account for this spread. Modelling the distribution of parallaxes of a population where all the stars are found at a same distance scattered by the parallax errors of the candidate members of California cloud results a much narrower distribution than the parallax spread of stars observed by Gaia identified as potential stellar members of the California stellar group (Figure \ref{fig:cali}, bottom left).

Recently, \cite{rezaei-kh.2022} have performed an investigation of the California cloud, reconstructing its 3-dimensional structure through using extinction of the field stars as a tracer of its depth. They similarly find that the cloud extends for $\sim$120 pc, and that its overall morphology is not a filament but a sheet. This is consistent with the distribution traced by the candidate members we see here.

The aforementioned hub containing the youngest stars appears to be located on the back side of the cloud (Figure \ref{fig:cali}, top right). There appears to be a gradient in tangential velocities as a function of distance of $\sim$2 \kms, with the gradient most apparent along the galactic longitude (Figure \ref{fig:cali}, bottom right). There may be a slight gradient in the radial velocities as well which would result in a differential expansion of the cloud, but the number of sources for which RVs are available is still relatively sparse, as such a larger number of stellar spectra are needed before it can be established.

\section{Discussion}\label{sec:discussion}

In this section we examine various processes that could have shaped the dynamics of Per OB2. We note that all three of them can coexist as they are applied over different spatial and temporal scales.

\subsection{Possibility of a past supernova}\label{sec:supernova}

Through clustering we have recovered 810 likely members towards Perseus, or, excluding the California Cloud, only 754 stars, down to the typical mass of 0.15-0.2 \msun \citep[masses were estimated by comparing the photometry relative to the MIST isochrones from][]{choi2016}. Even when accounting for more dispersed stars that cannot be recovered with clustering, this is approximately an order of magnitude fewer stars than its massive neighbor, the Orion Complex, which can boast membership in excess of $>10^4$ stars \citep{kounkel2019a}.

The bluest star in the catalog has [BP]-[RP] color of $\sim$0 mag, which roughly corresponds to late B -- early A type spectral class. However, Gaia may not be suitable for recovering most massive members of the nearby associations: among physical processes that may degrade their astrometry (such as, e.g., high degree of multiplicity), they may be susceptible to the bright limit of Gaia.

Previously, \citet{de-zeeuw1999} have conducted a survey of higher mass stars in OB associations with Hipparcos. They identified 17 OB type stars in Per OB2, of which three of the most massive ones are 40 Per \citep[B0.5V, found on the outskirts of Per OB2a][]{hoffleit1964} $\xi$ Per \citep[O7.5III,][]{sota2011}, as well as supergiant, $\zeta$ Per \citep[B1Ib, projected towards the densest part of Alcaeus][]{hoffleit1964}. Given that the maximum age of any group in Per OB2 is about 7 Myr, could there have existed a hitherto unknown star more massive than these two that would have had an opportunity to already reach an end of its life as a supernova?

While $\xi$ Per is a member of Per OB2, it is believed to be a runaway \citep{hoogerwerf2001}. With heliocentric radial velocity of 65 \kms\ \citep{gontcharov2006}, it is far in excess of the typical RV of $\sim$16--20 \kms\ typically observed in the region. Unfortunately, as $\xi$ Per does not currently have reported astrometry from Gaia, it is difficult to establish its precise distance. Hipparcos places it at $381^{+92}_{-62}$ pc \citep{van-leeuwen2007}. This is not dissimilar to the distances to some of the regions within Per OB2, such as 358 pc for Cynurus, 370 pc for Electryon, 390 pc for Mestor (the closest group to it on the plane of the sky), and 400 pc for Heleus. Even if it had originated in IC 348 (315 pc), given its large RV, at most it would have been ejected $<1-2$ Myr ago, consistent with the traceback estimate from \citep{hoogerwerf2001}. Using the mass and radius of 36 \msun\ and 14 \rsun\ for $\xi$ Per from \citet{krticka2010}, interpolated across MIST isochrones \citep{choi2016}, we estimate the age of the star of $\sim$3--4 Myr, which is larger than the traceback time.

There are a few processes that are capable of ejecting OB stars such as $\xi$ Per. One of the more famous of such massive runaways are AE Aur and $\mu$ Col, which have been ejected from the Orion Nebula due to dynamical interactions with other massive stars within the cluster \citep{hoogerwerf2001}. However, in case of $\xi$ Per, there is no ``matching pair'' of another star of comparable or greater mass, either in Per OB2a itself, or outside of it. Given that in the ejection scenario, it is typically the lower mass star that would undergo the greater acceleration, and given comparatively low densities even in the densest parts of Per OB2 (in comparison to the Orion Nebula), it seems unlikely that a currently accounted member of Per OB2 could be responsible for the ejection \citep{leigh2013,stone2019,manwadkar2020,manwadkar2021}. That is to say while a counterpart to a low mass runaway may be difficult to identify, it is extremely unlikely to eject the singularly most massive star in a given population. Thus, it could be theorized, that $\xi$ Per may have had a more massive companion. And, as $\xi$ Per has not travelled far, it is unlikely that this hypothetical companion would have had an opportunity to get far away from its birth site either.

Another method of producing a runaway is a supernova explosion. If a supernova goes off in a binary system, the system could lose substantial fraction of its mass and may become unbound. This will allow a companion to escape. This may be consistent with the scenario of $\xi$ Per. This suggests that a supernova explosion may have taken place in Per OB2. But, given the low mass of this association, its young age, and a relative scarcity of O-type stars within it -- even in the younger regions, it is doubtful that there have been several supernovae.

If the companion to $\xi$ Per existed and has since died, what implications would it have for Per OB2? Could it be responsible for the velocity patterns observed within it?

The neighboring Orion Complex has, to date, formed at least 4 stars - possibly more - that have produced a supernova \citep{kounkel2020a}. Two of them currently can be traced back as runaway pulsars, Geminga and PSR B0656+14 \citep{faherty2007,golden2005}, and two more can be inferred based on the effect they had on the rest of the Complex \citep{kounkel2018a,grossschedl2021}. In particular, they have triggered star formation, launching the newly formed stars on a ballistic trajectory away from the epicenter with speeds in excess of 6 \kms, producing a bubble that can be traced with stars several 10s of pc in diameter matching the accompanying bubbles seen in H$\alpha$, both $\lambda$ Ori and the Barnard's loop \citep{ochsendorf2015}. Furthermore, a supernova is likely responsible for the assembly of sufficient quantity of gas needed to form a cluster as massive as the Orion Nebula \citep{kounkel2021a}.

Comparing Per OB2 to the Orion Complex reveals a stark difference. While Per OB2a is expanding, it cannot be compared to the aforementioned ballistic expansion in Orion \citep[compare to Figure 2 in][]{kounkel2020a}. Similarly, Per OB2a and OB2b do not collectively form a bubble. The two oldest populations, Cynurus and Alcaeus would have formed in bulk prior to any possible supernova, and they already show two distinct velocity components of OB2a and OB2b. The younger generations of stars mostly inherit the velocity of these two older counterparts - no velocity gradient is found near them that would be consistent with triggering (Figure \ref{fig:per3d}).

Had a supernova occurred, it most likely happened after the bulk of the molecular gas in Per OB2a has already been expelled. As has been previously mentioned, since $\xi$ Per has been ejected $<$1-2 Myr ago, this is also the likely timeframe for the supernova explosion. Without molecular gas, the shockwave would not be able to interact with an already formed stellar population. The only exception to this may be IC 348. Assuming that the molecular cloud that can be seen towards it is associated with the cluster, it has held to its molecular gas for the longest period of time - longer than its neighbors, including those that have started forming stars at the same time, such as Heleus or Mestor (all three have age estimate within 0.5 Myr, Table \ref{tab:ave}). As a result, unlike them, it has been able to sustain active star formation since the initial burst of formation activity.

Comparing the fraction of dusty young stellar objects between IC 348, Mestor, and Heleus, using candidates identified in an all sky survey with WISE \citep[to keep similar sensitivity in both regions,][]{marton2016}, shows that despite some similarity in age, IC 348 has a much higher disk fraction, $\sim$10\%, versus $\sim$6\% for Mestor and for Heleus. However, the disk fraction for all three of them is much lower than NGC 1333, at $\sim$40\%, despite the same age derived for both NGC 1333 and IC 348 via isochrone fitting. That is to say, the shockwave may have played a role in stripping some stars of their disks in a region that is depleted of molecular gas.

While there has not been any obvious triggered star formation within the Perseus molecular cloud, there is some evidence of a shock applied onto the gas. As has been recently observed by \citet{tahani2022}, the magnetic fields towards the molecular cloud have a shape of an ark, which suggests a shock-cloud-interaction model. They establish that the shock has originated from behind the cloud, i.e., from Per OB2a applied onto Per OB2b. Similarly, they find that HI appears to have been swept towards us by the shockwave as well, with the ionized gas that is blueshifted relative to the cloud. Additionally, examining the three-dimensional structure of the molecular cloud from \citet{zucker2021} relative to the populations within Per OB2 (Figure \ref{fig:per3d}), the cloud has a pronounced bend as though it has been flattened out by an external force, with the bend being parallel to a spherical shockwave that may have originated in Per OB2a.

C$^{18}$O kinematics of the Perseus molecular cloud show a very complex velocity structure \citep{curtis2010}. It could be a response of the cloud to the shock.

\subsection{Possibility of cloud-cloud interaction} \label{sec:ccc}

As can be inferred in the data, Per OB2a and OB2b appear to have originated from two distinct clouds that both independently started forming stars when they were in a close proximity to one another. When the star formation has begun 5--7 Myr ago, corresponding to the age of the oldest groups in both regions, the ``hubs'' of star formation were some 65 pc apart (Figure \ref{fig:per3d}). However the true distance separating these two clouds is difficult to infer: star formation traces only the densest parts of a cloud, and the outer layers of a cloud could be more amorphous and diffuse. Following the initial burst of star formation, both of them continued to form stars in rapid succession, resulting in very similar ages for both populations.

It is difficult to say to what degree two populations have influenced each other, or if the two clouds would have started forming stars in a very similar manner if they were independent of one another \citep{dale2015}, but there does appear to be some degree of interaction between two populations.

The proximity of two clouds forming stars suggests a possibility of some type of cloud-cloud interaction. Future dynamical simulations would be able to constrain its exact nature, but a gravitational influence or a collision are a possibility. We note that while the shockwave from a supernova discussed in Section \ref{sec:supernova} would have been a recent development, occurring within the last 2 Myr, various types of interactions would have predated the supernova, shaping the dynamical evolution of the region from the very beginning. IC 348 stands out as a site where interactions between two clouds may have had the greatest influence, as it is caught in the middle, and it has similarity in its kinematics with both Per OB2a and OB2b. It is also worth noting that IC 348 is the most massive part of Per OB2 in terms of its stellar content, having more than twice as many stars than any other individual subgroup. The map from \citet{mcbride2021} also confirms IC 348 as the densest region in Per OB2, and based on the overall morphology of each region, it may be the only one to survive in the future as an open cluster. Being initially in a proximity to the molecular gas from both clouds (Per OB2a and OB2b) may be a reason why it has been able to reach this density in comparison to its neighbors.

While a collision of some kind is a possible type of interaction, typically, when cloud-cloud collisions are considered in simulations, the conditions are somewhat idealized. The clouds tend to be spherical, relatively small ($<$10 pc in size -- smaller than $\sim$50 pc scale corresponding to the current size of Per OB2b, which does not appear to significantly change its length when evolved back in time), moving towards each other head on, often featuring an impact of a larger cloud by a more compact bullet-like cloud \citep[e.g.,][]{takahira2014,haworth2015,fukui2021}. Even when more complex geometry is considered, or in cases where the scales are better approximation of the populations in Perseus \citep{wu2017,duarte-cabral2011}, the resulting collision is such that the two clouds merge completely. This has not occurred so far in Perseus, and a merger seems unlikely to be an eventual fate of Per OB2.

If Per OB2a and OB2b have experienced a collision, most likely it is only a grazing one, rather than a ``classic'' head-on, with two clouds influencing each other from further away. The self-gravity of the individual clouds may have assisted to find a point of contact (such as in IC 348), but that contact would have been limited outside of it. As previously mentioned, the current morphology of the Perseus molecular cloud suggests that it has been pushed away by a shock originating from Per OB2a. If the cloud originally has been straight, without a peculiar bend, then in the past this cloud would have been in a much greater proximity to IC 348 than what is currently observed. In such circumstances, a contact between the top of the cloud and IC 348 could have been likely, resulting in a collision.

Such a grazing collision is not typically explored in the simulations. Future modelling of the initial conditions in Per OB2 may better reveal the role that the two possible progenitor clouds have played in each other's evolution.

\subsection{Implications from older stars for the Per-Tau shell}

Recently, \citet{bialy2021} have examined a 3-dimensional dust map from \citet{leike2020}, and suggested a possibility of a spherical shell with diameter of $\sim$160 pc between Taurus and Perseus molecular clouds. They suggest that a previous supernova that has occurred 6--22 Myr ago may be responsible for this shell, and that in the process it has swept up the molecular gas to form Taurus and Perseus. A similar thought has been expressed by \citet{bally2008a} and \citet{chen2021} based on 2d studies.

We examine this claim using stellar counterparts to these molecular clouds. Taurus was not included in the clustering analysis presented in this paper, and, indeed, due to its diffuse nature it can be difficult to fully recover through clustering. As such, we adopt the average properties of the stellar groups identified by \citet{krolikowski2021}, and analyze their distribution and their kinematics with respect to those we see in Perseus.

While there is some degree of similarity between the kinematics of Per OB2b and Taurus, assuming the current velocities, at no point in the last 20 Myr would these two regions have been closer than 100 pc from each other, comparable to their present day separation. This does not change even if we consider Galactic potential through a traceback done in GalPy \citep{galpy}. That is to say, there is no apparent expansion between these two populations.

Indeed, \citet{bialy2021} do acknowledge that based on the CO velocities of the molecular gas, at the present day the shell is not expanding. They claim that the shell is sufficiently old to have fully decelerated. As not a trace of this supposed expansion that might have once took place remains, this may be difficult to conclusively prove one way or another with the currently available data.

However, more importantly, even if the lack of expansion can be disregarded, there is no obvious progenitor that may have caused such a ``bubble'' between Taurus and Perseus, which could have created the observed Per-Tau shell. Given the maximum age of Per OB2b of $\sim$5 Myr, it is too young to have hosted such a progenitor. As has been discussed in Section \ref{sec:supernova} - while there may have been a supernova in Perseus, it would have been relatively recent, it wouldn't have triggered significant star formation. 

Similarly, it is extremely unlikely for a progenitor to have originated in Taurus. Containing significantly fewer stars overall \citep[the most comprehensive membership list to date contains only 658 stars,][]{krolikowski2021}, it has only a few early type stars, with the most massive member being $\tau$ Tau \citep[B3V,][]{mooley2013}, which is 4 times less massive than $\xi$ Per, and would be expected have a lifespan of $\sim$100 Myr. Furthermore, consisting primarily of a group of diffuse cirrus-like clouds, it is highly unlikely for any star formation to have been triggered within Taurus by an external force, such as massive star feedback. The bulk of stars found in Taurus have an age $<$2 Myr. There is an older $\sim$16 Myr old population in the vicinity of the Taurus molecular clouds \citep{kraus2017,krolikowski2021}, but the stars within it are similarly diffuse, and similarly low mass, lacking early type stars.

Additionally, very few young stars are found between Taurus and Perseus. Examining maps of PMS stars produced by \citet{kounkel2019a}, \citet{kerr2021}, or \citet{mcbride2021}, there are few to no clustered groups with an age younger than 100 Myr found between 200 pc and the distance Perseus, which is where the progenitor would have to originate. While Taurus itself can be considered a ``low mass star forming region'', it is several times more massive than the total number of isolated young star candidates found in that volume of space. As such, a sufficiently massive star that could have produced a supernova is even less likely to originate in that space.

This does not prevent other possibilities, such as a type Ia supernova from an older field star, or a supernova from a massive runaway ending up far away from its parent association. However, without any dynamical signatures (obvious signatures of expansion), coordination in triggered star formation events between the clouds, or an obvious progenitor, a bubble scenario, to explain the Per-Tau shell, seems somewhat unlikely with the current evidence. The 3d cloud distribution presented by \citet{bialy2021} does not form a complete shell, and it can just as easily be interpreted as two independent sheet-like clouds. More future work is needed to understand the relation between Taurus and Perseus -- as well as other other star forming regions in the solar neighborhood. 

\citet{bialy2021} have presented other signals presumably originating from within Per-Tau complex, such as HI bubble or X-ray emission. It is difficult to understand how much weight to put into them. Compact high energy signatures are unlikely to persist for several Myr following a supernova \citep{ramakrishnan2020}, and, indeed, none are detected in a younger and far more obvious bubble in Orion \citep[See Figure 4 in][]{kounkel2020a}. HI bubble is far too small to be associated with Per-Tau shell, and, as has been concluded by \citet{tahani2022} from examining the magnetic topology, as well as the correlations between CO and HI velocities, HI is being blown from behind Perseus Molecular cloud, not from between Taurus and Perseus. H$\alpha$ traces primarily Per OB2 itself, not any structure outside of it. As such, these signatures are just as likely to have been caused by the recent activity within the older parts of Per OB2 than to be an independent confirmation of Per-Tau ``shell''. 

\section{Conclusions} \label{sec:concl}

We performed hierarchical clustering analysis in 5-dimensional phase space towards Per OB2 to identify 9 groups younger than 7 Myr. We estimate ages of these groups and perform a dynamical traceback to analyze their star forming history.

Similarly to \citet{pavlidou2021}, we find that the groups within Perseus tend to share two distinct systematic velocities that have no particular orientation relative to one another. IC 348 is caught in the middle, originating from Per OB2a, but it is moving towards Per OB2b, it may be in part affected by the gravity of the massive molecular cloud.

We further examine likely causes to such a configuration. It is possible that the two clouds have interacted with one another as the bulk of star formation has occurred within two distinctly moving clouds when they were in close proximity to one another. Some collision in vicinity of IC 348 may also have occurred.

There is a possibility of a past supernova that has originated within Per OB2a. Most likely it is relatively recent, with an age $<1-2$ Myr, with an age estimate corresponding to an ejection of a massive runaway $\xi$ Per. Per OB2b seem to have experience a shock from an impact \citep{tahani2022} that may be attributable to the supernova, with the shock traced by the morphology of the magnetic field lines, the relative velocities of CO and HII gas, as well as the overall structure of the cloud. However, comparing it to other theorized historical supernovae, e.g., in Orion \citep{kounkel2020a,grossschedl2021} or Vela \citep{cantat-gaudin2019b,pang2021}, Perseus does not show any obvious signatures of triggered star formation. Per OB2a does show slight signature of expansion, but that expansion could be attributed to the association being unbound, especially after gas dispersal -- it does not appear to be a ballistic expansion observed in other regions.

It has been theorized that there exists a shell between Taurus and Perseus molecular clouds, possibly from a bubble formed from another ancient supernova \citep{bialy2021}. We examine this claim, but are unable to support it with stellar kinematics. Similarly, we are unable to identify a likely site from which a progenitor could have originated. The age of Per OB2 precludes a possibility of anything being formed within it that would have caused such a bubble. There are no other young populations in the vicinity that have formed early type stars that would create a supernova in a reasonable time frame. While there may be a possibility of Type Ia supernova, or a massive runaway originating from much further away, this sheds some doubt on whether Perseus and Taurus molecular clouds are necessarily tracing a shell around a bubble, and requires more investigation in the future.

Per OB2 combined with the dynamical analysis of other star forming regions in the literature highlight disparate dynamical processes that may be involved in evolution of star forming regions. No two regions share the same history, and initial conditions greatly influence the structure and kinematics of each population. However, through performing a dynamical traceback, it is possible to uncover the dominant processes that have shaped each individual region. While currently there are only a few computational models that capture the full extent of complexity of these star forming regions, such tracebacks should be able to inform more complex simulations of these regions in the future.

\software{TOPCAT \citep{topcat}, Plotly \citep{plotly}}

\acknowledgments
We thank the anonymous referee for their detailed comments and suggestions. We also thank Mehrnoosh Tahani for the discussion of the magnetic field structure towards Perseus.
This work has made use of data from the European Space Agency (ESA)
mission {\it Gaia} (\url{https://www.cosmos.esa.int/gaia}), processed by
the {\it Gaia} Data Processing and Analysis Consortium (DPAC,
\url{https://www.cosmos.esa.int/web/gaia/dpac/consortium}). Funding
for the DPAC has been provided by national institutions, in particular
the institutions participating in the {\it Gaia} Multilateral Agreement.

\bibliographystyle{aasjournal.bst}
%\bibliography{references.bib}
\bibliography{main.bbl}

\end{document}